\documentclass[12pt]{article}
\usepackage{scicite}
\usepackage{graphicx}
\usepackage{placeins}
\usepackage{url}
\usepackage{lineno}
\usepackage[T1]{fontenc} % modern font encoding makes more symbols/operators work
\usepackage{gensymb} % for degree symbol

\usepackage{times}

\newenvironment{sciabstract}{%
\begin{quote} \bf}
{\end{quote}}

\usepackage{color}
\usepackage{xcolor}

\newcounter{lastnote}

\title{Ultra-low field \textsuperscript{13}C MRI of hyperpolarized pyruvate}

\author{Thomas Boele,$^{1,2\ast}$ Stephen J. McBride,$^{3}$ Megan Pike,$^{3}$\\ Erica Curran,$^{3}$ Patrick TomHon,$^{2,3,4}$\\ Hester Braaksma,$^{2,5}$ Sheng Shen,$^{2,6}$ Neha Koonjoo,$^{2,6}$\\ David E. Korenchan,$^{2,6}$, Eduard Chekmenev,$^{4}$ Thomas Theis,$^{3,7}$\\ David E. J. Waddington$^{1}$ and Matthew S. Rosen$^{2,6,8}$
	\vspace{0.5cm}\\
    \vspace{0.3cm}\parbox{1.0\linewidth}{\centering\normalsize{$^{1}$} Image X Institute, Sydney School of Health Sciences, Faculty of Medicine and Health, University of Sydney, NSW 2006, Australia}\\
    \vspace{0.2cm}\parbox{1.0\linewidth}{\centering\normalsize{$^{2}$} A. A. Martinos Center for Biomedical Imaging, Massachusetts General Hospital, 149 Thirteenth St., Charlestown, MA 02129, USA}\\
    \vspace{0.25cm}\parbox{1.0\linewidth}{\centering\normalsize{$^{3}$}  Department of Chemistry, North Carolina State University, Raleigh, NC 27606, United States}\\
    \vspace{0.35cm}\parbox{1.0\linewidth}{\centering\normalsize{$^{4}$} Integrative Biosciences, Department of Chemistry, Karmanos Cancer Institute, Wayne State University, Detroit, MI 48202, USA}\\
    \vspace{0.20cm}\parbox{1.0\linewidth}{\centering\normalsize{$^{5}$} Zernike Institute for Advanced Materials, University of Groningen, 9747 AG Groningen, The Netherlands}\\
	\parbox{1.0\linewidth}{\centering\normalsize{$^{6}$} Harvard Medical School, 25 Shattuck St., Boston, MA 02115, USA}\\
	\parbox{1.0\linewidth}{\centering\normalsize{$^{7}$} Department of Physics, North Carolina State University, Raleigh, NC 27606, USA}\\
    \parbox{1.0\linewidth}{\centering\normalsize{$^{8}$} Department of Physics, Harvard University, 17 Oxford St., Cambridge, MA 02138, USA}\\
	\normalsize{$^\ast$Corresponding author; Email: thomas.boele@sydney.edu.au}
}

\date{}
\begin{document} 


%\baselineskip24pt


\maketitle


%%%%%%%%%%%%%%%%%%%%%%%%%%%%%%%%%%%%%%%%%%%%%%%%%%%%%%%%%%%%%%%%%%%%%%%%%%%%%%%%%%%%%%%%%%%
%%%%%%%%%%%%%%%%%%%%%%%%%%%%%%%%%%%%%%%%%%%%%%%%%%%%%%%%%%%%%%%%%%%%%%%%%%%%%%%%%%%%%%%%%%%
%%%%%%%%%%%%%%%%%%%%%%%%%%%%%%%%%%%%%%%%%%%%%%%%%%%%%%%%%%%%%%%%%%%%%%%%%%%%%%%%%%%%%%%%%%%

\clearpage

\modulolinenumbers[5]
%\linenumbers

\begin{sciabstract}
%\textcolor{red}{150 word limit}
Medicine is evolving beyond therapy largely predicated on anatomical information and towards incorporating patient-specific molecular biomarkers of disease for more accurate diagnosis and effective treatment. The complementary combination of hyperpolarization by spin-lock induced crossing signal amplification by reversible exchange (SLIC SABRE) and low field magnetic resonance imaging (MRI) can enable accessible metabolic imaging to advance personalized medicine. Hyperpolarized $^{13}$C-enriched pyruvate has demonstrated utility for MRI of metabolism in cancer, heart disease and neurodegenerative disorders but has been restricted from widespread clinical adoption by a lack of access to affordable technology. Parahydrogen-based polarization techniques, paired with low-cost high-performance MRI at millitesla fields, offer a means of broadening the reach of metabolic imaging. Here we show results demonstrating in situ hyperpolarization of pyruvate at 6.5~mT by SLIC SABRE, followed by immediate readout without field cycling or sample shuttling. We achieve $^{13}$C signal enhancements several million times above thermal equilibrium at 6.5~mT, corresponding to polarization levels of approximately 3\%. Leveraging this enhancement, we perform $^{13}$C MRI and acquire NMR spectra with resolution sufficient to distinguish chemical shifts between pyruvate isotopomers. These results show a viable pathway towards accessible metabolic imaging with hyperpolarized $^{13}$C MRI at ultra-low field.

\end{sciabstract}

\section*{Introduction}
%%%%%%%%%%%%%%%%%%%%%%%%%%%%%%%%%%%%%%%%%%%%%%%%%%%%%%%%%%%%%%%%%%%%%%%%%%%%%%%%%%%%%%%%%%%
%%%%%%%%%%%%%%%%%%%%%%%%%%%%%%%%%%%%%%%%%%%%%%%%%%%%%%%%%%%%%%%%%%%%%%%%%%%%%%%%%%%%%%%%%%%
%%%%%%%%%%%%%%%%%%%%%%%%%%%%%%%%%%%%%%%%%%%%%%%%%%%%%%%%%%%%%%%%%%%%%%%%%%%%%%%%%%%%%%%%%%%

While changes in cellular metabolism are associated with many diseases, the majority of diagnostic imaging only measures macroscopic anatomical changes that occur after disease progression or treatment response. This paradigm is especially prevalent in cancer, where altered energy metabolism fuels unrestricted cell growth and change in tumor volume is the dominant measure of disease progression.\cite{Hanahan2011} In addition to helping clinicians better delineate tumor boundaries, metabolic imaging can aid cancer treatment by grading tumor malignancy and monitoring treatment response.\cite{Overcast2021} To enable agile personalized medicine there is a need for fast, accessible and non-ionizing metabolic imaging. 

Magnetic resonance imaging (MRI) is the premier imaging modality for soft tissue contrast with high spatial resolution, free from ionizing radiation. Beyond anatomical information, MRI also offers insight into brain activity, the microstructure of tissues and blood perfusion by providing functional, diffusion and dynamic contrast modalities. 

MRI can also perform molecular imaging via magnetic resonance spectroscopy (MRS) and chemical exchange saturation transfer (CEST). These techniques have shown promise for imaging metabolic biomarkers in cancer but the inherent sensitivity limits of MRI reliant on thermal Boltzmann polarization of nuclear spins have hampered widespread clinical adoption of these techniques, which are technically challenging to implement within clinically acceptable time frames.\cite{Galijasevic2022,Durmo2018,Jones2017,Cohen2022}

An emerging alternative is hyperpolarized MRI, which overcomes the sensitivity limits of MRI by boosting the nuclear polarization of a sample of spins many orders of magnitude above Boltzmann equilibrium. This process enables metabolic imaging within seconds following injection of a hyperpolarized metabolite, where the downstream metabolic products are detected in real time with chemical shift imaging. $^{13}$C MRI for metabolic imaging is in over 50 clinical trials for translation and has demonstrated utility for imaging metabolic hallmarks of disease, especially cancer.\cite{Kurhanewicz2019,Wang2019,Frijia2024,Jorgensen2022} Pyruvate is the most widely utilized metabolite, due to its central role in several metabolic pathways and its favorable relaxation properties. $^{13}$C pyruvate has shown promise as a hyperpolarized MRI contrast agent for measuring the metabolism of cancers of the prostate,\cite{Sushentsev2022,Nelson2013} pancreas,\cite{Stodkilde-Jorgensen2020} liver,\cite{Lee2021} kidney,\cite{Tran2019,Tang2021} breast,\cite{Gallagher2020,Woitek2021} and brain.\cite{Miloushev2018,Zaccagna2022}

However, clinical research with $^{13}$C MRI has been largely confined to only a few centers worldwide. The limited reach of this demonstrably useful method has been primarily due to challenges associated with the available hyperpolarization technology based on dynamic nuclear polarization (DNP).\cite{Ardenkjaer-Larsen2003} Polarizers using dissolution DNP typically hyperpolarize samples by transferring spin polarization from electron spins at magnetic fields of several Tesla and temperatures of a few Kelvin. This is followed by rapid dissolution and sample extraction by the injection of high-pressure superheated water into the polarizer where it can freeze, blocking or breaking the fluid path. These complications associated with the dissolution DNP method lead to slow throughput and high costs of DNP polarizers (1hr / sample from commercial polarizers costing >\$2M) and have restricted broad clinical adoption of $^{13}$C MRI.

Parahydrogen-induced polarization (PHIP) offers an affordable and fast alternative to DNP. Polarizers using PHIP methods can be built at a cost of ~\$10k and generate hyperpolarized samples within a few seconds.\cite{Schmidt2022,Ellermann2021,Nantogma2024} Spin order from the parahydrogen (pH$_2$) singlet state is used to boost the MRI signal of molecules either by direct hydrogenation of unsaturated precursors\cite{Bowers1986,Reineri2015} or via a reversible chemical reaction in a process called signal amplification by reversible exchange (SABRE).\cite{Adams2009} PHIP techniques lagged DNP technology for generating hyperpolarized $^{13}$C metabolites as reaching comparable $^{13}$C polarization levels in non-toxic metabolite solutions proved challenging. Preclinical polarizers based on PHIP side arm hydrogenation (PHIP SAH) are now commercially available \cite{Nagel2023} and clinical devices are on the horizon. The PHIP SAH approach involves costly and difficult development of individually optimized chemical reaction pathways between precursor-metabolite pairs. Consequently, pyruvate is the only metabolite currently available from this first generation of commercial PHIP SAH polarizers. In comparison, SABRE offers a more flexible process where the hydrogenation step is reversibly carried out with the aid of the SABRE catalyst. The reversible and repeatable SABRE process also offers opportunities outside single-shot delivery of biocompatible solutions, enabling experiments that benefit from many repetitions of the hyperpolarization step over time. Recent advances in SABRE and especially spin-lock induced crossing SABRE (SLIC SABRE) have demonstrated high polarization of $^{13}$C metabolites\cite{MacCulloch2023,Schmidt2023,Svyatova2021} and compatible methods for sample purification.\cite{Schmidt2022ACS,deMaissin2023,McBride2025} These breakthroughs position SLIC SABRE as a fast, flexible, accessible method poised to offer unprecedented access to $^{13}$C metabolic imaging. 

Combining these new breakthroughs in fast, accessible and low-footprint hyperpolarization with the complementary advantages of ultra-low field (ULF) MRI opens a pathway to broad adoption of metabolic imaging. ULF MRI is the perfect partner technology to SLIC SABRE for achieving the ultimate of broadening the impact of metabolic MRI. To date, the inaccessibility of metabolic MRI has been compounded by the drawbacks of high-field MRI where costs and exacting siting requirements restrict universal access. ULF MRI is a flexible cost-effective alternative that can be deployed in environments where high-field MRI is impossible.\cite{Sarracanie2015,Marques2019,Wald2020} ULF MRI has demonstrated utility in diagnosis of stroke,\cite{Mazurek2021,Yuen2022} hydrocephalus\cite{Obungoloch2018} and neurodegenerative disease\cite{Sorby-Adams2024} and has the potential to enable transformative portable screening paradigms.\cite{Shen2025preprint,Shen2025T1map,Mallikourti2024}

The limitation of ULF MRI is the inherently low sensitivity of inductive detection of weakly polarized spins. Hyperpolarization methods decouple spin polarization from thermal equilibrium at the detection field and offer a way to overcome the SNR limits of ULF MRI.\cite{Sarracanie2015} Hyperpolarized $^{13}$C offers the additional advantage of background-free contrast. A further motivation for pursuing SABRE hyperpolarization at low field is that it can be combined with fields matching conditions for the generation of spin polarization. For example, 6.5~mT is the ideal field to perform $^1$H-SABRE hyperpolarization of molecules such as pyridine and pyrazine.\cite{Adams2009,Iqbal2024} Adding $^{13}$C SLIC SABRE hyperpolarization expands the tool chest, uniquely positioning our platform at 6.5~mT as a test bed for exploring hyperpolarization dynamics in the mT regime. 

Here we show rapid hyperpolarization of $^{13}$C-enriched pyruvate in an ULF open-access MRI scanner. Within 10~s, SLIC SABRE generates enhancements in $^{13}$C polarization of pyruvate at 6.5~mT of approximately 6 orders of magnitude. The resulting high-resolution spectra are sufficiently narrow to detect chemical shift and J-coupling effects and reveal features indicating both catalyst-bound and free hyperpolarized pyruvate species. We characterize the build-up and $T_1$ decay dynamics of SLIC~SABRE at 6.5~mT for pyruvate and show the optimum spin-lock parameters for generating transverse $^{13}$C magnetization on pyruvate from singlet order. Leveraging this optimization we demonstrate $^{13}$C MRI in two modes -- a single shot sequence utilizing one hyperpolarization step compatible with an in vivo imaging paradigm and a multiple shot sequence for interrogating emergent polarizer dynamics. 

These results demonstrate the flexibility and utility of SLIC~SABRE, especially for MRI sequence development when deployed as a hyperpolarization technique that can repeatedly re-polarize a sample within the sensing region for iterative optimization. Our results motivate future in vivo experiments performing $^{13}$C MRI at ultra-low field and have potential to enable accessible metabolic imaging to accelerate the transition towards personalized medicine.

\FloatBarrier

%%%%%%%%%%%%%%%%%%%%%%%%%%%%%%%%%%%%%%%%%%%%%%%%%%%%%%%%%%%%%%%%%%%%%%%%%%%%%%%%%%%%%%%%%%%
%%%%%%%%%%%%%%%%%%%%%%%%%%%%%%%%%%%%%%%%%%%%%%%%%%%%%%%%%%%%%%%%%%%%%%%%%%%%%%%%%%%%%%%%%%%
%%%%%%%%%%%%%%%%%%%%%%%%%%%%%%%%%%%%%%%%%%%%%%%%%%%%%%%%%%%%%%%%%%%%%%%%%%%%%%%%%%%%%%%%%%%
\section*{Results}
%%%%%%%%%%%%%%%%%%%%%%%%%%%%%%%%%%%%%%%%%%%%%%%%%%%%%%%%%%%%%%%%%%%%%%%%%%%%%%%%%%%%%%%%%%%
%%%%%%%%%%%%%%%%%%%%%%%%%%%%%%%%%%%%%%%%%%%%%%%%%%%%%%%%%%%%%%%%%%%%%%%%%%%%%%%%%%%%%%%%%%%
%%%%%%%%%%%%%%%%%%%%%%%%%%%%%%%%%%%%%%%%%%%%%%%%%%%%%%%%%%%%%%%%%%%%%%%%%%%%%%%%%%%%%%%%%%%

%%%%%%%%%%%%%%%%%%%%%%%%%%%%%%%%%%%%%%%%%%%%%%%%%%%%%%%%%%%%%%%%%%%%%%%%%%%%%%%%%%%%%%%%%%%
%%%%%%%%%%%%%%%%%%%%%%%%%%%%%%%%%%%%%%%%%%%%%%%%%%%%%%%%%%%%%%%%%%%%%%%%%%%%%%%%%%%%%%%%%%%
\subsection*{SLIC SABRE enhances \textsuperscript{13}C signal by > 10\textsuperscript{6} $\times$ at ULF}
%%%%%%%%%%%%%%%%%%%%%%%%%%%%%%%%%%%%%%%%%%%%%%%%%%%%%%%%%%%%%%%%%%%%%%%%%%%%%%%%%%%%%%%%%%%
%%%%%%%%%%%%%%%%%%%%%%%%%%%%%%%%%%%%%%%%%%%%%%%%%%%%%%%%%%%%%%%%%%%%%%%%%%%%%%%%%%%%%%%%%%%

%%%%%%%%%%%    FIGURE 1    %%%%%%%%%%%%
\begin{figure} %%% "figure*" for double-column width figure? %%%
\begin{center}
\includegraphics[width=16cm]{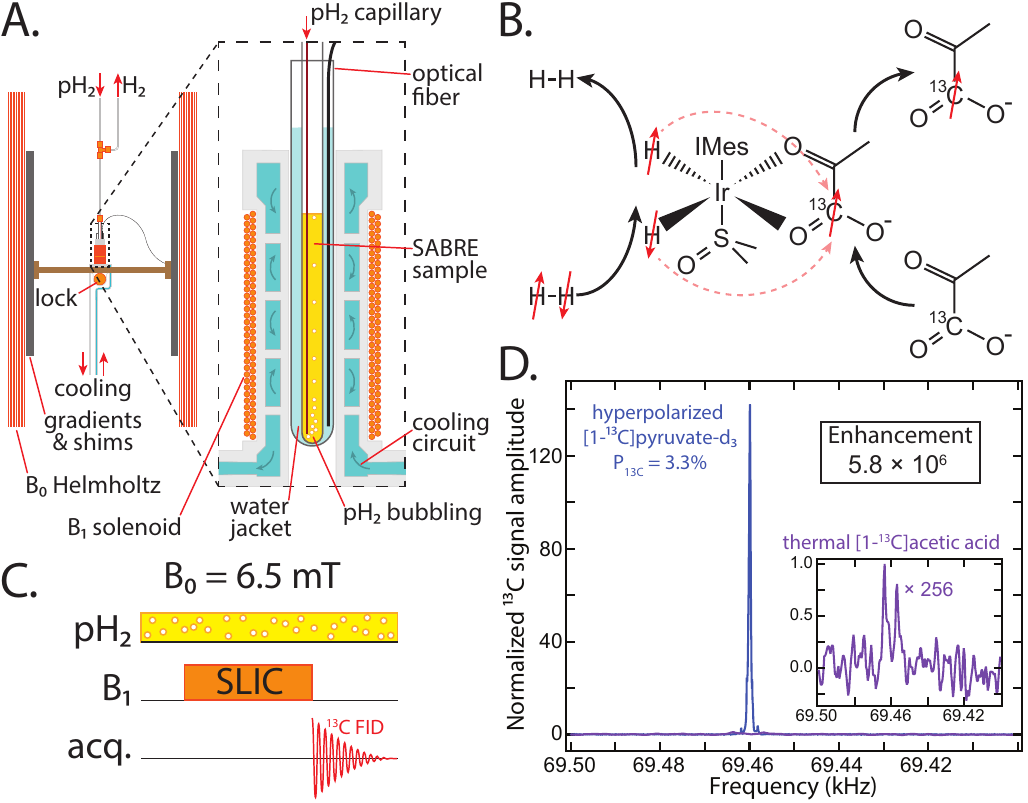} %17.4cm original image
\caption{\textbf{SLIC SABRE hyperpolarizes pyruvate at 6.5~mT.} \textbf{A.} Experimental setup schematic, showing a front-on view of the inside of the 6.5 mT MRI scanner and an expanded cross-section of the variable-temperature 69.5~kHz NMR probe. Temperature and pH$_2$ flow are controlled by cooling water and hydrogen gas circuits that extend outside the RF-shielded scan room. \textbf{B.} Chemical exchange between the Ir-based organometallic complex and free pH$_2$ and pyruvate in solution facilitates bulk hyperpolarization. Dashed red arrows indicate the spin-lock driven transfer of spin order. \textbf{C.} Pulse sequence diagram of a generalized SLIC SABRE experiment. During steady pH$_2$ bubbling a spin-lock pulse is applied on resonance with $^{13}$C at an amplitude corresponding to an effective field close to the J-coupling between the hydrides bound to the Ir complex. Magnetization builds up in the transverse plane along the vector of the spin-lock in the frame rotating at the $^{13}$C Larmor frequency. Following the spin lock, magnetization is immediately detected as a free-induction decay. \textbf{D.} Spectra of hyperpolarized deuterated [1-$^{13}$C]pyruvate and thermal [1-$^{13}$C]acetic acid. The pyruvate spectrum was acquired in a single 32~s acquisition following 120~s SLIC SABRE with 24 sccm pH$_2$ flow through a sample containing 11.4~µmol pyruvate. Insert shows the [1-$^{13}$C]acetic acid spectrum acquired with 256 averages of a 45\degree~pulse every 15~s applied to a 31.4~mmol sample.}
\label{Fig_Setup}
\end{center}
\end{figure} 
%%%%%%%%%%%%%%%%%%%%%%%%%%%%%%%%%%%%
We begin our results showing hyperpolarized pyruvate enhancement above thermal equilibrium at 69.5~kHz, alongside diagrams of the SABRE reaction and experimental setup, in Figure~\ref{Fig_Setup}. The reversible chemical exchange reaction that underpins the SLIC SABRE results shown here is depicted in Figure~\ref{Fig_Setup}B. At its heart is the iridium-based SABRE catalyst that acts to reversibly bind both hydrogen and pyruvate, facilitating the transfer of spin order from pH$_2$ to $^{13}$C driven by SLIC. Bulk hyperpolarization is achieved through a repeated cycle of fresh pH$_2$ binding, transfer of spin order to unpolarized pyruvate, and the release of the orthohydrogen and spin-polarized pyruvate. For this cycle to result in significant polarization of $^{13}$C pyruvate free in solution, there are four broad requirements: Firstly, a supply of parahydrogen to fuel hyperpolarization and to displace orthohydrogen by continuous bubbling. Secondly, appropriate concentrations of precatalyst, stabilizing ligand and substrate in solution. Thirdly, correct radiofrequency (RF) spin-lock frequency and amplitude. Fourthly, and finally, favorable chemical exchange rates for both pH$_2$ and pyruvate. Appropriate sample composition has been extensively characterized previously \cite{Tickner2020} and we selected a solution composition of pyruvate, DMSO and SABRE precatalyst in deuterated methanol in suitable concentrations determined by previous work, described in Materials and Methods below. Our experimental setup, providing control over the other 3 requirements - pH$_2$ delivery, sample temperature and RF - is shown in Figure~\ref{Fig_Setup}A and described in the Methods Section. Briefly, pH$_2$ is bubbled through the SABRE sample inside the 6.5~mT scanner via a capillary that ends in the bottom of the sample tube. The pH$_2$ flow rate is controlled by a gas handling circuit outside the RF-shielded enclosure of the scanner. RF excitation is performed by a solenoid coil wrapped around a hollow former, through which coolant flows to control the temperature of the SABRE solution.

The general sequence for a SLIC SABRE experiment is shown in Figure~\ref{Fig_Setup}C. Following catalyst activation, with the sample cooled to 4.4\degree C and pH$_2$ bubbling continuously, the application of a spin-lock pulse at the $^{13}$C frequency with an amplitude corresponding to the J-coupling between the hydride groups and $^{13}$C nucleus of the bound pyruvate substrate causes net magnetization to build up in the transverse plane, aligned with the spin-lock vector in the frame rotating at the $^{13}$C Larmor frequency. An example frequency spectrum of SLIC SABRE hyperpolarized [1-$^{13}$C]pyruvate-d$_3$ is plotted in Figure~\ref{Fig_Setup}D, alongside an inset showing the thermal signal from a [1-$^{13}$C]acetic acid sample. 

The signal enhancement here is an estimate calculated by comparison to a larger, highly concentrated $^{13}$C-enriched acetic acid sample, described in detail in Supplementary Section 1. This enhancement represents an increase in $^{13}$C polarization of $5.8\times10^6$ for deuterated pyruvate. This corresponds to $^{13}$C polarization of approximately 3.3\%. While this $^{13}$C polarization is lower than has been achieved with SABRE under other conditions, here we employed a low pH$_2$ flow rate to minimize solvent boil-off and displacement from the NMR coil, allowing for experiments with a single SABRE sample over several hours. Hyperpolarization at higher pH$_2$ flow rate is anticipated to yield $^{13}$C polarization values in excess of 10\%.\cite{Schmidt2022ACS,deMaissin2023,McBride2025}

\FloatBarrier
%%%%%%%%%%%%%%%%%%%%%%%%%%%%%%%%%%%%%%%%%%%%%%%%%%%%%%%%%%%%%%%%%%%%%%%%%%%%%%%%%%%%%%%%%%%
%%%%%%%%%%%%%%%%%%%%%%%%%%%%%%%%%%%%%%%%%%%%%%%%%%%%%%%%%%%%%%%%%%%%%%%%%%%%%%%%%%%%%%%%%%%
\subsection*{Pyruvate $^{13}$C imaging with SLIC-shot MRI}
%%%%%%%%%%%%%%%%%%%%%%%%%%%%%%%%%%%%%%%%%%%%%%%%%%%%%%%%%%%%%%%%%%%%%%%%%%%%%%%%%%%%%%%%%%%
%%%%%%%%%%%%%%%%%%%%%%%%%%%%%%%%%%%%%%%%%%%%%%%%%%%%%%%%%%%%%%%%%%%%%%%%%%%%%%%%%%%%%%%%%%%
%%%%%   SLIC-shot MRI Figure    %%%%%
\begin{figure}
\begin{center}
\includegraphics[width=16cm]{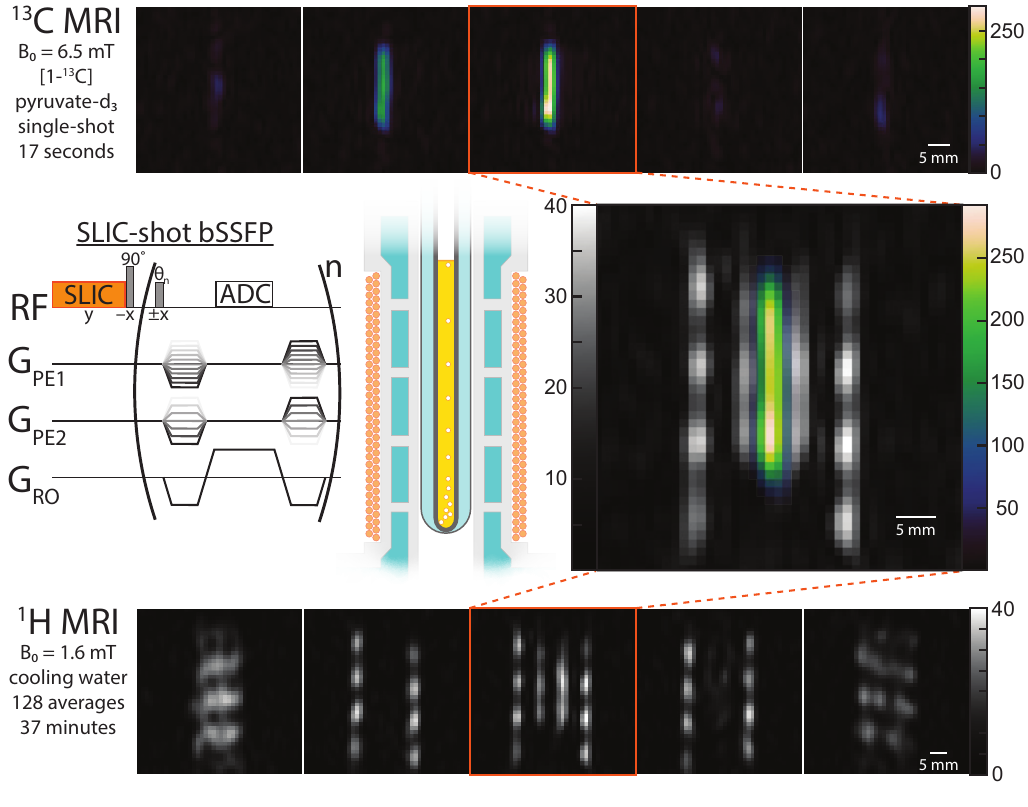}
\caption{\textbf{SLIC-shot $^{13}$C MRI at 6.5~mT.} Top row: 512$\times$64$\times$5 3D $^{13}$C MRI of [1-$^{13}$C]pyruvate-d$_3$ following a 120~s SLIC pulse with 24 sccm pH$_2$ bubbling and 50\% undersampling. Middle row: Left, MRI sequence using a single SLIC-shot to generate $^{13}$C hyperpolarization, followed by a bSSFP-like sequence with a variable tip angle to sample the hyperpolarized signal in a near-steady fashion. Center, the region inside the solenoid of the variable temperature NMR probe in schematic cross section. Right, Co-registered $^{13}$C and $^1$H 3D MRI images, with the $^1$H image a single slice of 3D bSSFP at 1.6~mT and the $^{13}$C image from above overlaid. Bottom row: 3D bSSFP of $^1$H performed by ramping $B_0$ down to 1.6~mT to bring protons in the cooling water of the variable temperature probe on resonance at 69.5~kHz. All MRI data was zero-filled by a factor of 2 in the readout and first phase encode directions and normalized in units of the noise floor.}
\label{Fig_SLICshotMRI}
\end{center}
\end{figure}
%%%%%%%%%%%%%%%%%%%%%%%%%%%%%%%%%%%%%
Having achieved in situ hyperpolarization of pyruvate at 6.5~mT we move to show results leveraging SLIC SABRE induced hyperpolarization for $^{13}$C MRI of pyruvate at 6.5~mT. Of the pyruvate molecules investigated, [1-$^{13}$C]pyruvate-d$_3$  was identified as the best for exploring $^{13}$C MRI, as it has the longest $T_1$ and reaches the highest in situ SLIC SABRE polarization (see Supplementary Sections 2, 3 and 5 for SLIC SABRE optimization and spin relaxation details). Envisaging future in vivo metabolic imaging, we begin by demonstrating $^{13}$C MRI in a single-shot mode in Figure \ref{Fig_SLICshotMRI}, where hyperpolarization is generated at the beginning of the experiment and stored for sampling later. The imaging sequence we developed for our ULF scanner is an adaptation of the 3D balanced steady-state free precession (bSSFP) sequence we have previously tailored to maximize SNR and contrast at ULF.\cite{Sarracanie2015,Waddington2020} The sequence diagram in Figure \ref{Fig_SLICshotMRI} highlights the addition of a SLIC SABRE hyperpolarization step that precedes the repeated bSSFP spatial encoding block. To preserve the transverse magnetization generated by the spin-lock, a 90\degree~pulse with orthogonal phase is applied immediately following SLIC to store magnetization along the longitudinal axis. To sample that stored magnetization steadily, a series of pulses of increasing tip angle follows, each of which tips part of the remaining magnetization back into the transverse plane.\cite{Deppe2012} We term this method SLIC-shot bSSFP, noting that the sequence is more pedantically "bSSFP-like", with steady state magnetization a goal requiring careful sampling of the remaining available magnetization rather than simply reached after many pulses with a freely precessing magnetization vector as is more typical in bSSFP.

The $^{13}$C images in Figure \ref{Fig_SLICshotMRI} were acquired with SLIC-shot bSSFP in 17~s following 120~s of SLIC SABRE hyperpolarization. To make experiments reliably repeatable, and as there is relatively little magnetic susceptibility distortion at ULF, pH$_2$ bubbling was maintained throughout at 24~sccm. This comes at the cost of minor motion artifacts, most pronounced in the second phase encode direction. The pyruvate sample is approximately 12~µmol in 400~µL of methanol. The scale bar is estimated based on the 3.8~mm inner diameter of the 5~mm NMR tube. 

To provide context and create an image in analogy to an anatomical background we also show a $^1$H image of the variable temperature NMR probe cooling circuit acquired by ramping $B_0$ down to 1.6~mT to bring protons on resonance at 69.5~kHz. We coregister the central slices from the $^{13}$C and $^1$H MRI data to show the hyperpolarized pyruvate solution in place in a 5mm high pressure NMR tube, inside the 10mm water-filled tube in thermal contact with the coolant-filled variable temperature NMR probe body.

%%%%%   multiSLIC-shot MRI Figure    %%%%%
\begin{figure}
\begin{center}
\includegraphics[width=16cm]{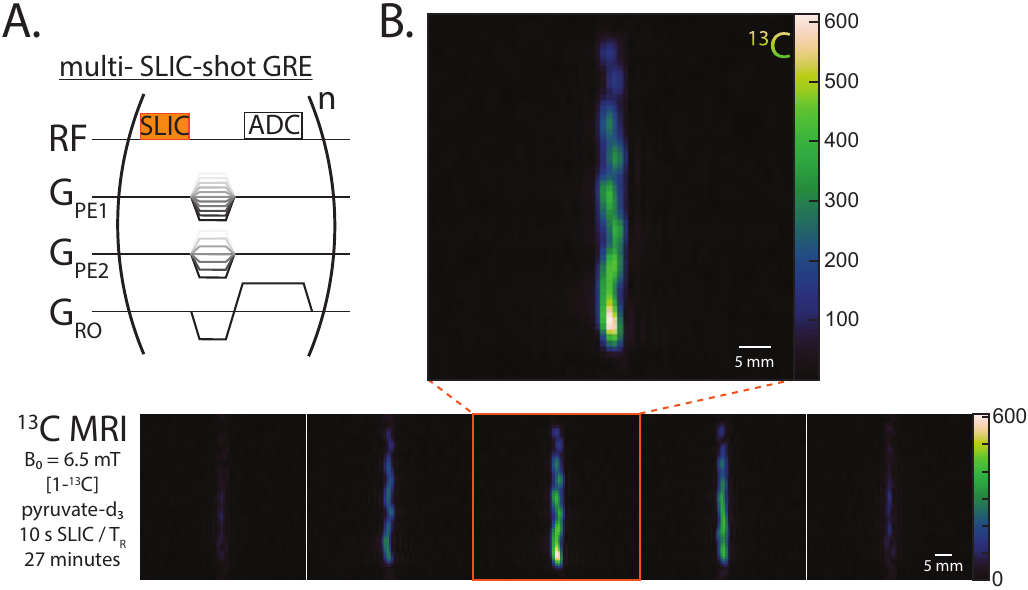}
\caption{\textbf{multiple SLIC-shot $^{13}$C MRI at 6.5mT. A.} MRI sequence for multiple SLIC-shot 3D gradient echo. A SLIC pulse is applied to generate fresh hyperpolarized signal for every line of k-space. \textbf{B.} 576$\times$64$\times$5 3D $^{13}$C MRI of [1-$^{13}$C]pyruvate-d$_3$ with the multi-SLIC-shot sequence in A. and 50\% undersampling, 10~s SLIC per shot and 24 sccm pH$_2$ bubbling. As the parahydrogen bubbling creates a constantly renewing helical pattern of bubbles in the tube, that shape is captured in the $^{13}$C MRI image despite the 27-minute total imaging time. All MRI data was zero-filled by a factor of 2 in the readout and first phase encode directions and normalized in units of the noise floor.}
\label{Fig_multiSLICshotMRI}
\end{center}
\end{figure}
%%%%%%%%%%%%%%%%%%%%%%%%%%%%%%%%%%%%%

While in vivo metabolic imaging requires fast imaging sequences that efficiently sample the available non-renewable magnetization, the design of polarizer and reactor technology able to produce larger volumes of hyperpolarized material efficiently can also benefit from slower imaging sequences that reveal emergent dynamics. With this secondary focus, we developed a gradient echo (GRE) sequence leveraging multiple spin lock hyperpolarization steps as shown in Figure \ref{Fig_multiSLICshotMRI}. The multi-SLIC-shot MRI sequence incorporates a SLIC pulse for each phase encoding step within a 3D GRE sequence, drawn in Figure \ref{Fig_multiSLICshotMRI}A. We note that there are no hard pulses in this imaging sequence as the spin-lock generates transverse magnetization directly. This initializes each line of k-space encoding with a steady $^{13}$C magnetization, eliminating artifacts from underlying signal variation without the need for a variable tip angle approach or k-space filtering. Example imaging results from the multi-SLIC-shot sequence, displayed in Figure \ref{Fig_multiSLICshotMRI}B, capture the standing pattern of hydrogen bubbling at a flow rate of 24~sccm. As with the single SLIC-shot MRI results, there are motion artifacts apparent, especially in the 2nd phase encode direction. However, characteristics of the in situ polarization method can be identified that show non-uniform polarization within the NMR tube and sample being forced out of the detection coil, highlighting opportunities to further improve hyperpolarization and sensitivity.

\FloatBarrier
%%%%%%%%%%%%%%%%%%%%%%%%%%%%%%%%%%%%%%%%%%%%%%%%%%%%%%%%%%%%%%%%%%%%%%%%%%%%%%%%%%%%%%%%%%%
%%%%%%%%%%%%%%%%%%%%%%%%%%%%%%%%%%%%%%%%%%%%%%%%%%%%%%%%%%%%%%%%%%%%%%%%%%%%%%%%%%%%%%%%%%%
\subsection*{High resolution spectroscopy resolves J-coupling and chemical shift at ULF}
%%%%%%%%%%%%%%%%%%%%%%%%%%%%%%%%%%%%%%%%%%%%%%%%%%%%%%%%%%%%%%%%%%%%%%%%%%%%%%%%%%%%%%%%%%%
%%%%%%%%%%%%%%%%%%%%%%%%%%%%%%%%%%%%%%%%%%%%%%%%%%%%%%%%%%%%%%%%%%%%%%%%%%%%%%%%%%%%%%%%%%%
%%%%%%% HIGH RESOLUTION SPECTRA FIGURE %%%%%%%
\begin{figure}
\begin{center}
\includegraphics[width=7cm]{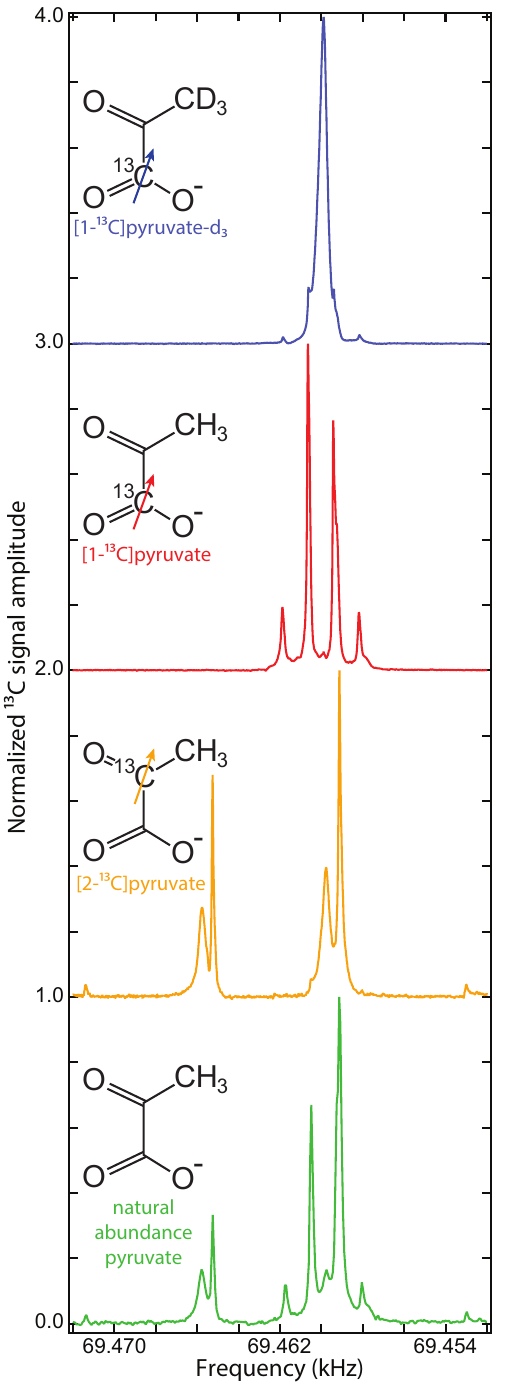} %8.6cm file
\caption{\textbf{SLIC SABRE enabled NMR spectroscopy at 6.5~mT.} $^{13}$C spectra of [1-$^{13}$C]pyruvate-d$_3$ (blue), [1-$^{13}$C]pyruvate (red), [2-$^{13}$C]pyruvate (yellow) and natural abundance pyruvate (green). Splittings are dominated by J-coupling to the protons on the methyl group creating a quartet of peaks for each $^{13}$C site. All $^{13}$C-enriched spectra were acquired in a single shot following 120 s of SLIC. The natural abundance spectrum was acquired with 64 averages of 20~s SLIC. SLIC frequency and amplitude were selected to maximize the integrated signal for each spectrum. All spectra were acquired at 24 sccm pH$_2$ and 4.4 ± 0.2\degree C and are normalized and offset for clarity.}
\label{Fig_Spectra}
\end{center}
\end{figure}
%%%%%%%%%%%%%%%%%%%%%%%%%%%%%%%%%%%%%%%%%%%%%%
Looking forwards from our proof of principle $^{13}$C MRI demonstration towards accessible metabolic imaging at ULF, we present high resolution pyruvate spectra acquired at 6.5~mT in Figure~\ref{Fig_Spectra}. While our $^{13}$C MRI results showcase an important step towards hyperpolarized ULF $^{13}$C metabolic imaging, in vivo studies tracking metabolism will require a means of distinguishing downstream products after injection. Conventionally in the few-Tesla magnetic field regime chemical shift readily separates the metabolic products of pyruvate from each other in frequency. In comparison, as chemical shift is proportional to $B_0$, the splittings in spectra acquired at ULF are dominated by J-coupling. For example, the chemical shift separation between [1-$^{13}$C]pyruvate and [1-$^{13}$C]lactate is 12.2~ppm, corresponding to 0.9~Hz at our field. However, the hyperpolarized $^{13}$C spectra shown in Figure \ref{Fig_Spectra} resolve chemical shift between pyruvate species and highlight the opportunities for CSI and spectroscopy in the mT regime.

Deuterated [1-$^{13}$C]pyruvate has the simplest spectrum, with a single primary peak with a full width at half maximum (FWHM) of 0.46~Hz. The fine features to either side of the primary peak are from a small fraction of pyruvate with a protonated methyl group. Protonated pyruvate exhibits a quartet of peaks associated with J-coupling to the methyl group protons. This splitting is absent from the deuterated pyruvate spectrum, where the J-coupling to the spin-1 deuterons is much weaker. The FWHM of the individual sharp peaks in the protonated spectrum is 0.16~Hz. We attribute the fine details in our [1-$^{13}$C]pyruvate spectra to imperfect enrichment and/or exchange of protons and deuterons occurring over time.\cite{Barskiy2014} To investigate other potential explanations we performed additional NMR spectroscopy at 11.7~T on SABRE solutions after measurement at ULF, detailed in Supplementary section 4.

The [2-$^{13}$C]pyruvate spectrum has the same quartet of peaks as [1-$^{13}$C]pyruvate, split further apart by stronger J-coupling between the $^{13}$C and methyl protons. We attribute the additional broad features to catalyst-bound pyruvate. These features are most apparent when inspecting the inner peaks of the [2-$^{13}$C]pyruvate spectrum but we think it likely features associated with bound pyruvate are present and unresolved in other spectra. The bound pyruvate peaks are broader and chemically shifted down field from the peaks we attribute to free [2-$^{13}$C]pyruvate, whereas bound [1-$^{13}$C]pyruvate is expected to shift up field, and to a lesser degree.\cite{Schmidt2023} The peaks associated with free pyruvate are narrowed by comparatively rapid molecular tumbling. The distinction is clearer to see on [2-$^{13}$C]pyruvate signal as the chemical shift difference between bound/free pyruvate is greater, stronger J-coupling helps separate all features and there is a lower fraction of free hyperpolarized pyruvate. We note the separation between the center of [1-$^{13}$C]pyruvate and [2-$^{13}$C]pyruvate spectra is from the different chemical shift experienced at these carbon positions.

The spectrum of natural abundance pyruvate helps to intuit the separate contributions from individual $^{13}$C spins that occur even in small impurities in the enriched spectra. A convolution of the [1-$^{13}$C]pyruvate and [2-$^{13}$C]pyruvate spectra accounts for all the features present in the spectrum from the unenriched sample, with the relative weighting of each contribution a function the ratio of the enhancements of each species after 20~s of SLIC SABRE. To account for the nearly one hundred times lower $^{13}$C signal in the natural abundance sample, signal averaging was performed and a shorter acquisition duration employed to reduce the overall experimental time to a reasonable 30 minutes. Although our primary focus here is upon understanding the hyperpolarization mechanisms and characteristics of pyruvate spectra at ULF in the context of metabolic imaging, we take the opportunity to suggest that the ability to collect J-coupling resolved spectra of natural abundance $^{13}$C molecules at ULF with µmol quantities of material could be highly advantageous for applications in chemical screening.

\FloatBarrier
%%%%%%%%%%%%%%%%%%%%%%%%%%%%%%%%%%%%%%%%%%%%%%%%%%%%%%%%%%%%%%%%%%%%%%%%%%%%%%%%%%%%%%%%%%%
%%%%%%%%%%%%%%%%%%%%%%%%%%%%%%%%%%%%%%%%%%%%%%%%%%%%%%%%%%%%%%%%%%%%%%%%%%%%%%%%%%%%%%%%%%%
%%%%%%%%%%%%%%%%%%%%%%%%%%%%%%%%%%%%%%%%%%%%%%%%%%%%%%%%%%%%%%%%%%%%%%%%%%%%%%%%%%%%%%%%%%%
\section*{Discussion}
Our results have demonstrated the utility of ULF MRI as a test-bed for SABRE $^{13}$C hyperpolarization experiments, broadened our understanding of hyperpolarization mechanisms and optimal conditions for the mT regime, shown pulse sequence development leveraging $^{13}$C signal enhancements $>10^{6}$ for 3D MRI of pyruvate, and reported chemical shift resolution and spin-lattice relaxation times compatible with spectroscopic imaging in vivo. Taken altogether, these findings demonstrate the clinical potential of ULF $^{13}$C MRI for the first time. We now discuss the advantages of the mT regime, the challenges and opportunities facing imaging metabolism at ULF and the necessary steps ahead for clinical translation.

%%%%%%%%%%%%%%%%%%%%%%%%%%%%%%%%%%%%%%%%%%%%%%%%%%%%%%%%%%%%%%%%%%%%%%%%%%%%%%%%%%%%%%%%%%%
%%%%%%%%%%%%%%%%%%%%%%%%%%%%%%%%%%%%%%%%%%%%%%%%%%%%%%%%%%%%%%%%%%%%%%%%%%%%%%%%%%%%%%%%%%%
%%%%%%%%%%%%%%%%%%%%%%%%%%%%%%%%%%%%%%%%%%%%%%%%%%%%%%%%%%%%%%%%%%%%%%%%%%%%%%%%%%%%%%%%%%%
\subsection*{Advantages of SLIC SABRE in the mT regime}
%%%%%%%%%%%%%%%%%%%%%%%%%%%%%%%%%%%%%%%%%%%%%%%%%%%%%%%%%%%%%%%%%%%%%%%%%%%%%%%%%%%%%%%%%%%
%%%%%%%%%%%%%%%%%%%%%%%%%%%%%%%%%%%%%%%%%%%%%%%%%%%%%%%%%%%%%%%%%%%%%%%%%%%%%%%%%%%%%%%%%%%
While previous results have shown the flexibility of SLIC SABRE by applying the method in the microtesla\cite{Ortmeier2024,Kempf2024} and tesla\cite{Pravdivtsev2015} regimes, our results show there are clear advantages to operating at millitesla magnetic fields.

Compared to SLIC SABRE performed at high magnetic fields, the millitesla regime has three key advantages: Firstly, pH$_2$ bubbling can be applied continuously without magnetic susceptibility effects causing distortion in the $B_0$ and $B_1$ fields. Secondly, there is no singlet-to-triplet mixing of parahydrogen.\cite{Berner2019} Thirdly, chemical shift effects are small enough that the singlet state between hydride groups is maintained without spin locking and the $^{13}$C Larmor frequency for bound and free pyruvate species overlap sufficiently for the SLIC pulse to spin-lock free pyruvate spins in the transverse plane following hyperpolarization. Overcoming these effects requires the additional complications of RF pulses to spin-lock protons and selective RF pulses to excite free and bound spins.\cite{DeVience2012,Theis2014,Pravdivtsev2014} These challenges have limited the pursuit of in situ SLIC SABRE hyperpolarization for $^{13}$C MRI at high-field, despite the potential of in-situ setups to provide fast, low-cost hyperpolarization platforms (avoiding the need for a separate polarizer device) and help visualize reaction dynamics to evaluate performance and inform reactor design. These benefits of in-situ hyperpolarized MRI have spurred PHIP experiments\cite{Schmidt2017,Schmidt2018,Mohiuddin2024} which offer a viewpoint from which to reflect upon our work. The hyperpolarized $^{13}$C MRI achieved to date with this in-situ method imaged a pyruvate PHIP-SAH precursor molecule inside a 7~T preclinical scanner.\cite{Mohiuddin2024} Our results compare favorably, for although preclinical gradients used in Ref\cite{Mohiuddin2024} enable sub-second experiments with sub-mm in-plane resolution, image signal to noise is comparable, if not inferior to our SLIC-shot MRI results despite our approximately ten-fold lower polarization starting point.

Comparing SLIC SABRE at millitesla fields to microtesla experiments, the crucial advantage is that conventional inductive detection is still practical and there is no need to resort to complicated superconducting magnetometer setups for readout.\cite{Buckenmaier2018} As a consequence, compared to the best recent pyruvate $^{13}$C MRI in the µT regime, our SLIC-shot MRI results achieve comparable resolution and SNR within less than 1\% of the imaging time (including both the hyperpolarization and MRI acquisition steps).\cite{Kempf2024} Further, 6.5~mT in particular is the ideal field to perform SABRE hyperpolarization of $^1$H on molecules such as pyridine and pyrazine, as this field strength is close to matching the level anti-crossing condition. This opens up opportunities for hyperpolarized imaging schemes at 6.5~mT that leverage hyperpolarization across multiple nuclei in an agile and flexible platform.\cite{Adams2009,Iqbal2024} Indeed, work from our laboratory and others' has already demonstrated hyperpolarization of spins beyond $^1$H and $^{13}$C is feasible and has utility for in vivo applications.\cite{Lehmkuhl2020,MacCulloch2023,Skre2025} 

Further work will also leverage the advantages present in our system. High homogeneity in $B_0$ and $B_1$ combine to make an excellent platform for exploring novel hyperpolarization schemes and transfer of polarization between nuclei. While the elegant simplicity of spin-lock pulses for generating hyperpolarization has driven the discoveries presented in this manuscript, the modest polarization reached in these initial results suggests there is significant potential for improvement. We hope to utilize the flexibility of our research system's hardware and pulse sequence programming to implement non-intuitive schemes for polarization transfer that do not rely on matching spin energy levels to anti-crossings.\cite{Eriksson2022} The favorable characteristics of our system also position us to explore generating long-lived spin states on $^{13}$C tracers that can be read out via proton magnetization for longer effective metabolic imaging timescales and greater sensitivity.\cite{Mandzhieva2022,Mandzhieva2023}

\FloatBarrier
%%%%%%%%%%%%%%%%%%%%%%%%%%%%%%%%%%%%%%%%%%%%%%%%%%%%%%%%%%%%%%%%%%%%%%%%%%%%%%%%%%%%%%%%%%%
%%%%%%%%%%%%%%%%%%%%%%%%%%%%%%%%%%%%%%%%%%%%%%%%%%%%%%%%%%%%%%%%%%%%%%%%%%%%%%%%%%%%%%%%%%%
\subsection*{Challenges and Opportunities for ULF $^{13}$C MRI}
%%%%%%%%%%%%%%%%%%%%%%%%%%%%%%%%%%%%%%%%%%%%%%%%%%%%%%%%%%%%%%%%%%%%%%%%%%%%%%%%%%%%%%%%%%%
%%%%%%%%%%%%%%%%%%%%%%%%%%%%%%%%%%%%%%%%%%%%%%%%%%%%%%%%%%%%%%%%%%%%%%%%%%%%%%%%%%%%%%%%%%%
Leaving aside the advantages of in-situ hyperpolarization in the ULF regime, the imaging platform we have showcased here is entirely compatible with a more traditional in vivo imaging paradigm where a polarizer operates outside the scanner and hyperpolarized samples are shuttled inside for imaging. In time further optimization of in situ SLIC SABRE may yield polarization levels comparable with those achieved by polarizers operating at other field strengths, but we envisage the speediest path to in vivo work at ULF will incorporate an external polarizer into an experimental method that uses ULF for imaging alone. This will address the need for larger boluses of polarized $^{13}$C metabolites and incorporate depressurization and purification steps. With those hurdles largely overcome by recent advances in SABRE technology,\cite{Schmidt2022ACS,deMaissin2023,McBride2025} in this subsection we turn to the challenges and opportunities specific to ULF $^{13}$C MRI. 

%MRI sequence development  
While a lack of chemical shift is advantageous for in situ SLIC SABRE hyperpolarization, it is less convenient when considering the challenges faced performing chemical shift imaging at ULF.\cite{Rodriguez2025} For example, it will preclude the use of interleaved spectrally selective 3D bSSFP sequences.\cite{deMaissin2023,Nagel2023} However, the spectra we have presented demonstrate that while challenging, CSI at 6.5~mT is by no means impossible. The $^{13}$C linewidths we have reported here are sufficiently narrow to encourage the pursuit of chemical shift imaging at 6.5~mT, despite reduced chemical shift in absolute frequency terms at low magnetic field. The 0.46~Hz linewidth we observe in deuterated pyruvate corresponds to 6.6~ppm at 69.5~kHz, which suggests that resolving the chemical shift of 12.2~ppm between [1-$^{13}$C]pyruvate and [1-$^{13}$C]lactate, while challenging, is achievable. More encouraging, resolving the 136.8~ppm separation between [2-$^{13}$C]pyruvate and [2-$^{13}$C]lactate is clearly practicable. It is necessary to acknowledge that the linewidths reported here are measured within a 5~mm NMR tube rather than across a several-cm volume of interest in vivo. However, our shims can be further optimized, and our spectra are broadened by the contribution from catalyst-bound pyruvate which will not be present in purified samples. Operating at low field also offers the opportunity to narrow $^{13}$C spectra of protonated molecules by employing proton decoupling pulses in vivo, as the higher specific absorbance rates that forestall such methods at T fields are near negligible in the mT regime.

Both MRI sequence and hardware development will be necessary to fulfill the potential of ULF metabolic MRI. To improve temporal resolution, we will accelerate MRI acquisition and increase gradient performance. For the former, the implementation of methods honed at high field, such as compressed sensing and deep learning, offers a pathway for future work so long as the specific noise characteristics at low field are taken into account.\cite{Efrat} For the latter, increased gradient efficiency is readily achievable for future in vivo applications by upgrading our hardware, applying methods already demonstrated for efficient gradient and shim design.\cite{Shen2022} Gradient strength is particularly critical to improve resolution for a low gamma nucleus at low field, and a factor of 4 improvement upon the 1~mT/m gradients used for our proof of principle $^{13}$C MRI here is feasible, especially for a preclinical demonstration. Long term however, the optimum choice of field strength for ULF metabolic imaging will be informed by the maximum gradient strength and field homogeneity achievable at an accessible cost. Other hardware development will include a more robust field lock, able to operate during MRI scanning, and dedicated $^1$H/$^{13}$C coils to remove the need for the impractical field cycling method that we utilized in this work to produce co-registered $^1$H/$^{13}$C images.

A challenge for all hyperpolarized MRI, regardless of field strength, is $T_1$ relaxation. While the highest 45~s $T_1$ value for pyruvate measured in our SABRE solutions at ULF is shorter than pyruvate $T_1$ times measured at high field, our $T_1$ is still sufficiently long to probe pyruvate-lactate conversion if replicated in vivo. Further, we expect purified pyruvate solutions to exhibit longer $T_1$ times, particularly if D$_2$O is utilized as the solvent.\cite{deMaissin2023} There are also opportunities to extend effective $T_1$ at ULF with long-lived singlet states. This takes advantage of the strong coupling regime at ULF to enable long-lived singlet states to persist until read out through the application of a spin-lock or trains of pulses practicable in vivo at ULF due to lower SAR and high field homogeneity.\cite{Mandzhieva2022} Beyond extending the effective time over which hyperpolarized metabolites maintain spin-order in vivo, schemes leveraging long-lived singlets offer readout methods that can distinguish pyruvate and lactate based on their different $^{13}$C J-couplings, and proton-only sensing that can further broaden the reach of hyperpolarized metabolic imaging by alleviating the need for specialized $^{13}$C hardware.\cite{Mandzhieva2023} 

In conclusion, parahydrogen-driven hyperpolarization techniques are uniquely positioned to take the extensive body of work from the dissolution DNP community on metabolic imaging and amplify its impact by making the modality widely accessible in the clinic. Leveraging this technology alongside ULF MRI presents an excellent opportunity for each technology to complement the other: hyperpolarization can bypass the SNR challenges inherent to Boltzmann polarization at low field, and presents contrast mechanisms in a regime where methods developed for high-field MRI don't work. ULF MRI can simultaneously increase the reach of metabolic imaging to clinical paradigms including population screening and remote locations. While challenges associated with spin relaxation and resolving subtle chemical shifts at low field are present, there is no fundamental limitation barring this technology. SLIC SABRE offers a fast, flexible and accessible method for hyperpolarizing $^{13}$C metabolites under a range of conditions to accelerate the translation of this technology from physics advance, through preclinical tool toward clinical application. In the ULF regime, where affordable and portable MRI scanners operate, there is significant potential to enable next generation molecular imaging with point of care scanners.

%%%%%%%%%%%%%%%%%%%%%%%%%%%%%%%%%%%%%%%%%%%%%%%%%%%%%%%%%%%%%%%%%%%%%%%%%%%%%%%%%%%%%%%%%%%
%%%%%%%%%%%%%%%%%%%%%%%%%%%%%%%%%%%%%%%%%%%%%%%%%%%%%%%%%%%%%%%%%%%%%%%%%%%%%%%%%%%%%%%%%%%
%%%%%%%%%%%%%%%%%%%%%%%%%%%%%%%%%%%%%%%%%%%%%%%%%%%%%%%%%%%%%%%%%%%%%%%%%%%%%%%%%%%%%%%%%%%
\section*{Materials and Methods}
%%%%%%%%%%%%%%%%%%%%%%%%%%%%%%%%%%%%%%%%%%%%%%%%%%%%%%%%%%%%%%%%%%%%%%%%%%%%%%%%%%%%%%%%%%%
%%%%%%%%%%%%%%%%%%%%%%%%%%%%%%%%%%%%%%%%%%%%%%%%%%%%%%%%%%%%%%%%%%%%%%%%%%%%%%%%%%%%%%%%%%%
%%%%%%%%%%%%%%%%%%%%%%%%%%%%%%%%%%%%%%%%%%%%%%%%%%%%%%%%%%%%%%%%%%%%%%%%%%%%%%%%%%%%%%%%%%%

%%% Preparation of SABRE sample %%%
\subsection*{Sample preparation}
Solutions for SABRE were prepared with 30~mM of either [1-\textsuperscript{13}C] or [1-\textsuperscript{13}C]-d$_3$ pyruvate, 6~mM  [IrCl(COD)(IMes)] precatalyst and 20~mM DMSO in 450~µL of methanol-d\textsubscript{4}. DMSO is added to modulate the exchange rate of the pyruvate as previously described.\cite{Iali2019,Tickner2020} Precatalyst was synthesized following established methods.\cite{Iali2019,Vazquez-Serrano2006,Cowley2011} Chemicals were purchased from Sigma-Aldrich and stored refrigerated under argon. All liquids were degassed with argon to displace dissolved paramagnetic oxygen, which otherwise acts to relax pH$_2$ to H$_2$. SABRE solutions were sonicated to ensure all precatalyst was dissolved. 

%%% pH2 generation %%%
\subsection*{pH$_2$ generation and delivery}
pH$_2$ was generated by flowing hydrogen gas over iron oxide hydroxide at 27~K at a rate of 2~slm and stored in aluminium cylinders at 480~psi before use within 24 hours. The pH$_2$ generator system is described in Supplementary Section 5.

%%% pH2 delivery %%%
To run a SLIC SABRE experiment, a SABRE solution was placed in a 5~mm high-pressure NMR tube (Wilmad) at 6.5~mT and bubbled with pH$_2$ at a pressure of 95~psi and rate controlled by the combination of a mass flow controller (MFC, Alicat) and back pressure regulator. The pH$_2$ delivery circuit, including a detailed schematic, is shown in Supplementary Section 5. A period of tens of minutes' pH$_2$ bubbling facilitates activation of the catalyst precursor by hydrogenation of the COD ligand. 12~sccm was found to be a reliable rate of pH$_2$ delivery for the majority of experiments, providing excellent enhancement whilst limiting sample evaporation.

%%% cooling circuit %%%
\subsection*{Temperature control}
To modulate the chemical exchange rate of pyruvate binding and releasing the SABRE catalyst, the 5~mm sample tube was cooled by being placed in a 10~mm water-filled NMR tube in thermal contact with the walls of a temperature-controlled 10~mm inner diameter NMR probe. The NMR probe was home-built using a 3D-printed hollow resin former, through which automotive coolant was pumped by a circulating chiller (Huber). The chiller was situated outside the RF-shielded room of the MRI scanner and connected to the temperature-controlled probe with 1/4" insulated tubing. The cooling circuit is described in Supplementary Section 5. Temperature was monitored with a fibreoptic thermometer (Osensa) placed in the 10~mm tube of water to account for the offset between the chiller setpoint and temperature at the sample inside the warm MRI scanner. 

%%% NMR hardware %%%
\subsection*{NMR hardware}
All measurements were performed in a custom-built MRI scanner with a Tecmag Redstone console described previously\cite{Sarracanie2015}. This scanner consists of a biplanar electromagnet, operating at 6.5~mT magnetic field, powered by a direct current (DC) supply (DANFYSIK 854~T) and biplanar magnetic gradient coils. Active shimming is achieved by applying DC offsets to the gradient coils via gradient amplifiers (AE Techron 7794MRL). The high $B_0$ homogeneity and field-locked stability of this scanner are particularly advantageous for spin-lock experiments.\cite{DeVience2013,DeVience2021} A $B_0$ field-frequency lock maintains proton resonance frequency within ±0.25~Hz and the scanner is shimmed to achieve a linewidth of deionized water of better than 0.5~Hz. The work described here used a 2-layer solenoid coil, wound about the 10~mm inner diameter 3D-printed hollow former using 5/39/42-AWG20 Litz wire with fluorinated ethylene propylene insulation (New England Wire Technologies). The coil was tuned to 69.5~kHz with an external resonator board in parallel-tune, series-match configuration. The tuned and matched coil had a q-factor of 22, corresponding to a bandwidth of 3~kHz, determined by an S21 measurement about 69.5~kHz with a vector network analyzer and an untuned pick-up coil weakly coupled to the NMR probe. To achieve the low power and stability needed for spin-lock pulses of a few µT $B_1$ amplitude, the power amplifier was bypassed and RF pulses used directly from the console synthesizer. To maintain the fidelity of the low power SLIC pulses, measurements employed an active CMOS-based transmit/receive switch (Mini-Circuits) with low power loss and low phase distortion.

% Typical SLIC experiment
Following catalyst activation and sample cooling, a typical SLIC SABRE experiment, shown in Figure~\ref{Fig_Setup}C, involved the application of a spin-lock pulse followed by immediate FID acquisition, with pH$_2$ bubbling constant throughout. Reliable and repeatable SLIC parameters on resonance at 69.5~kHz were 10~s spin-lock pulse duration, 10~Hz spin-lock amplitude and 12~sccm pH$_2$ flow rate. Experiments exploring individual parameters -- detailed in Supplementary Sections 2 and 5 -- used these values as a starting point for controlling variables not under investigation. The free induction decay acquired following the spin-lock pulse was converted to a frequency spectrum via fast Fourier transform and phase corrected to bring the real part of the signal purely into the absorption mode. The real part of the spectrum was then integrated between ±50Hz. The spectra shown in Figures \ref{Fig_Setup} and \ref{Fig_Spectra} were zero filled by a factor of 2.

\subsection*{MRI}
Imaging was carried out with the same variable temperature NMR probe as the spectroscopy measurements. For $^{13}$C MRI sequences the field lock was applied during SLIC SABRE but blanked while imaging gradients were applied. 

The SLIC-shot $^{13}$C MRI sequence is a 3D bSSFP sequence adapted with the addition of spin-lock and 90\degree pulses prefacing a bSSFP gradient block with a variable tip angle. The SLIC-shot MRI sequence used an echo time ($T_E$) of 54~ms, a repetition time ($T_R$) of 108~ms and a 512$\times$64$\times$5 matrix undersampled by 50\%. The variable tip schedule was calculated following the iterative method described in Reference \cite{Deppe2012} with an initial tip angle of 1\degree. Total duration for acquiring the 160 lines of k-space pulse was 17.3~s.

$^1$H MRI was performed by ramping the electromagnet down to 1.6~mT to bring protons on resonance at 69.5~kHz to match the NMR probe tuning. The $^1$H MRI used a bSSFP sequence with the same $T_E$, $T_R$ and undersampled matrix as the $^{13}$C SLIC-shot MRI, a tip angle of 45\degree, and 128 averages for a total imaging time of 37 minutes. The field lock was not used at 1.6~mT as it was far outside its frequency bandwidth. Shim values were not re-optimized for 1.6~mT.

The multi-SLIC-shot $^{13}$C MRI sequence was adapted from a gradient echo sequence with the substitution of a spin-lock pulse for the standard hard pulse. The multi-SLIC-shot scan used $T_E=$87~ms, $T_R=$121~ms and a 10~s spin-lock for each line of 50\% undersampled 576$\times$64$\times$5 matrix. This equates to a total sequence duration of 27 minutes.

Image reconstruction applied DC offset and ADC turn-on spike corrections, zero filled k-space by a factor of 2 in the readout and first phase encode directions and performed a 3D inverse Fourier transform. Normalization in units of the noise floor was performed by dividing the signal in image space by the standard deviation of a region of interest comprising the bottom left 1/8 of the FOV.

%%%%%%%%%%%%%%%%%%%%%%%%%%%%%%%%%%%%%%%%%%%%%%%%%%%%%%%%%%%%%%%%%%%%%%%%%%%%%%%%%%%%%%%%%%%
%%%%%%%%%%%%%%%%%%%%%%%%%%%%%%%%%%%%%%%%%%%%%%%%%%%%%%%%%%%%%%%%%%%%%%%%%%%%%%%%%%%%%%%%%%%
%%%%%%%%%%%%%%%%%%%%%%%%%%%%%%%%%%%%%%%%%%%%%%%%%%%%%%%%%%%%%%%%%%%%%%%%%%%%%%%%%%%%%%%%%%%
\bibliographystyle{sciadv}
\bibliography{scibib2024}
%%%%%%%%%%%%%%%%%%%%%%%%%%%%%%%%%%%%%%%%%%%%%%%%%%%%%%%%%%%%%%%%%%%%%%%%%%%%%%%%%%%%%%%%%%%
%%%%%%%%%%%%%%%%%%%%%%%%%%%%%%%%%%%%%%%%%%%%%%%%%%%%%%%%%%%%%%%%%%%%%%%%%%%%%%%%%%%%%%%%%%%
%%%%%%%%%%%%%%%%%%%%%%%%%%%%%%%%%%%%%%%%%%%%%%%%%%%%%%%%%%%%%%%%%%%%%%%%%%%%%%%%%%%%%%%%%%%

%%%%%%%%%%%%%%%%%%%%%%%%%%%%%%%%%%%%%%%%%%%%%%%%%%%%%%%%%%%%%%%%%%%%%%%%%%%%%%%%%%%%%%%%%%%
\subsection*{Acknowledgements}
%%%%%%%%%%%%%%%%%%%%%%%%%%%%%%%%%%%%%%%%%%%%%%%%%%%%%%%%%%%%%%%%%%%%%%%%%%%%%%%%%%%%%%%%%%%

%%%%%%%%%%%%%%%%%%%%%%%%%%%%%%%%%%%%%%%%%%%%%%%%%%%%%%%%%%%%%%%%%%%%%%%%%%%%%%%%%%%%%%%%%%%
\textbf{General:} The authors thank Stephen E Ogier for sharing the computer assisted design file of the hollow coil former used to build the variable temperature probe and Earl Emery for conceiving and carrying out custom modification to the Redstone receiver for additional gain below 100~kHz.

%%%%%%%%%%%%%%%%%%%%%%%%%%%%%%%%%%%%%%%%%%%%%%%%%%%%%%%%%%%%%%%%%%%%%%%%%%%%%%%%%%%%%%%%%%%
\textbf{Funding:}  This work was supported by the U.S. National Institutes of Health, Research Project grant 1R01EB034197-01A1, the U.S. National Institute of Biomedical Imaging and Bioengineering under award number R21EB033872, the U.S. Department of Energy, Office of Biological and Environmental Research, under award numbers DE-SC0023334 and DE-SC0025315, and the U.S. National Science Foundation under grant CHE-2404387. The views and opinions of authors expressed herein do not necessarily state or reflect those of the United States Government or any agency thereof. D.E.J.W. and T.B. received support from Australian National Health and Medical Research Council Investigator Grants 2017140 and 1194004, respectively. T.B. acknowledges the support of a Fulbright Future Scholarship, funded by The Kinghorn Foundation, from the Australian–American Fulbright Commission. MSR acknowledges the generous support of the Kiyomi and Ed Baird MGH Research Scholar award.

%%%%%%%%%%%%%%%%%%%%%%%%%%%%%%%%%%%%%%%%%%%%%%%%%%%%%%%%%%%%%%%%%%%%%%%%%%%%%%%%%%%%%%%%%%%
\textbf{Author contributions:} 
Conceptualization: TB, TT and MSR.
Methodology: TB, SJM, MP, EC, PT, SS, NK, EC, TT and MSR.
Investigation: TB, HB, NK, DK.
Visualization: TB and DEJW.
Funding acquisition: TT, EC and MSR. %TB Fulbright small fry. DEJW as stand in for PK?
Project administration: DEJW, TT, MSR.
Supervision: TB, TT, DEJW and MSR.
Writing – original draft: TB.
Writing – review \& editing: All.

%%%%%%%%%%%%%%%%%%%%%%%%%%%%%%%%%%%%%%%%%%%%%%%%%%%%%%%%%%%%%%%%%%%%%%%%%%%%%%%%%%%%%%%%%%%
\textbf{Data availability:} The authors declare that the data supporting the findings of this study are available within the article and its supplementary information. Raw data are available from the corresponding author upon reasonable request.

%%%%%%%%%%%%%%%%%%%%%%%%%%%%%%%%%%%%%%%%%%%%%%%%%%%%%%%%%%%%%%%%%%%%%%%%%%%%%%%%%%%%%%%%%%%
\textbf{Competing Interests:} M.S.R. is a founder and equity holder of Hyperfine Inc. M.S.R. is an equity holder of DeepSpin GmbH. M.S.R. also serves on the scientific advisory boards of ABQMR, Synex Medical, Nanalysis, and O2M Technologies. E.Y.C. is a co- founder and equity holder of XeUS Technologies LTD.

\clearpage

\setcounter{figure}{0}
\renewcommand{\figurename}{Supplementary Figure}    % changes the text before each figure

\title{Supplementary Material -- Ultra-low field \textsuperscript{13}C MRI of hyperpolarized pyruvate}
% Ultra-low field $^{13}$C MRI of pyruvate hyperpolarized in situ via SLIC SABRE
% $^{13}$C MRI of hyperpolarized pyruvate at 6.5 mT 
% Ultra-low field SLIC-shot $^{13}$C MRI of SABRE hyperpolarized pyruvate


%\baselineskip24pt


\maketitle


%%%%%%%%%%%%%%%%%%%%%%%%%%%%%%%%%%%%%%%%%%%%%%%%%%%%%%%%%%%%%%%%%%%%%%%%%%%%%%%%%%%%%%%%%%%
%%%%%%%%%%%%%%%%%%%%%%%%%%%%%%%%%%%%%%%%%%%%%%%%%%%%%%%%%%%%%%%%%%%%%%%%%%%%%%%%%%%%%%%%%%%
%%%%%%%%%%%%%%%%%%%%%%%%%%%%%%%%%%%%%%%%%%%%%%%%%%%%%%%%%%%%%%%%%%%%%%%%%%%%%%%%%%%%%%%%%%%

\clearpage

\modulolinenumbers[5]
\linenumbers

%%%%%%%%%%%%%%%%%%%%%%%%%%%%%%%%%%%%%%%%%%%%%%%%%%%%%%%%%%%%%%%%%%%%%%%%%%%%%%%%%%%%%%%%%%%
%%%%%%%%%%%%%%%%%%%%%%%%%%%%%%%%%%%%%%%%%%%%%%%%%%%%%%%%%%%%%%%%%%%%%%%%%%%%%%%%%%%%%%%%%%%
%%%%%%%%%%%%%%%%%%%%%%%%%%%%%%%%%%%%%%%%%%%%%%%%%%%%%%%%%%%%%%%%%%%%%%%%%%%%%%%%%%%%%%%%%%%
\section*{Supplementary Section 1: $^{13}$C polarization estimation}
%%%%%%%%%%%%%%%%%%%%%%%%%%%%%%%%%%%%%%%%%%%%%%%%%%%%%%%%%%%%%%%%%%%%%%%%%%%%%%%%%%%%%%%%%%%
%%%%%%%%%%%%%%%%%%%%%%%%%%%%%%%%%%%%%%%%%%%%%%%%%%%%%%%%%%%%%%%%%%%%%%%%%%%%%%%%%%%%%%%%%%%
%%%%%%%%%%%%%%%%%%%%%%%%%%%%%%%%%%%%%%%%%%%%%%%%%%%%%%%%%%%%%%%%%%%%%%%%%%%%%%%%%%%%%%%%%%%
Thermal polarization of $^{13}$C at 6.5~mT is too low to measure enhancement directly on 30~mM of $^{13}$C enriched pyruvate so an estimation was performed by comparing the signal from thermal 1-$^{13}$C enriched acetic acid as shown in Supplementary Figure \ref{Fig_13CPolCal}. Acetic acid was selected because it is readily available in neat liquid form and not dissimilar in composition to pyruvate. The acetic acid spectrum is split by the 6.8~Hz J-coupling between the $^{13}$C and methyl protons.\cite{Kupriyanov2021} Despite measuring for more than an hour, our signal-to-noise was only sufficient to confidently identify 2 peaks of the anticipated quartet. Knowing the molar concentration of $^{13}$C in a 1.8~mL sample of neat 1-$^{13}$C acetic acid, the gyromagnetic ratio, thermal Boltzmann polarization, and applying the Ernst equation with an estimated acetic acid T$_1$ of 40~s we calculated the enhancement in signal from hyperpolarized pyruvate compared to acetic acid to give the 3.3\% estimate of $^{13}$C polarization quoted in the main text. Matlab code used to calculate the estimate follows:

{\scriptsize
\begin{verbatim}
signal_SABRE = 9.3907e8;
signal_ace = 4.8494e7;
theta = 45; % tip angle in degrees
TR = 15;    % repetition time in s
T1 = 40;    % estimated T1 in s
mass_pyruvate_2mL = 0.0065;     %mass of pyruvate in g in 2mL
mass_pyruvate = mass_pyruvate_2mL/2*0.4;    % " " in 400 µL
M_mass_pyruvate = 114.04;    % molar mass [1-13C]pyruvate-d3
M_pyruvate = mass_pyruvate/M_mass_pyruvate;     % Moles of pyruvate
M_mass_ace = 61.04;
density_ace = 1.066;
M_ace = 1.8*density_ace/M_mass_ace;     % Moles of acetic acid
molar_ratio = M_ace/M_pyruvate;

% calculate steady-state Mz after repeated pulses
M_zSS =  (1-exp(-TR/T1))/(1-cosd(theta)*exp(-TR/T1));
M_xySS = sind(theta)*M_zSS;
signal_ace_thermal90 = signal_ace/256 * (1/M_xySS);
signal_ace_thermal90_M = signal_ace_thermal90/M_ace;
signal_pyr_M = signal_SABRE/M_pyruvate;

B=6.5e-3;   % Magnetic field strength in Tesla
T = 300;    % Temperature in K
gyro_13C=10.705e6;  %gyromagnetic ratio for 13C in Hz/T/2pi
hbar=1.05457173e-34;    %Planck's const. h/(2*pi)
k_B=1.3806488e-23;  %Boltzmann's const.
thermal_P_13C=tanh((gyro_13C*hbar*2*pi*B)./(2*k_B*T))*100;

enhancement = signal_pyr_M / signal_ace_thermal90_M
13C_polarization = enhancement*thermal_P_13C
\end{verbatim}
}

%%%%%   13C POLARIZATION CALCULATION FIGURE    %%%%%
\begin{figure}
\begin{center}
\includegraphics[width=16cm]{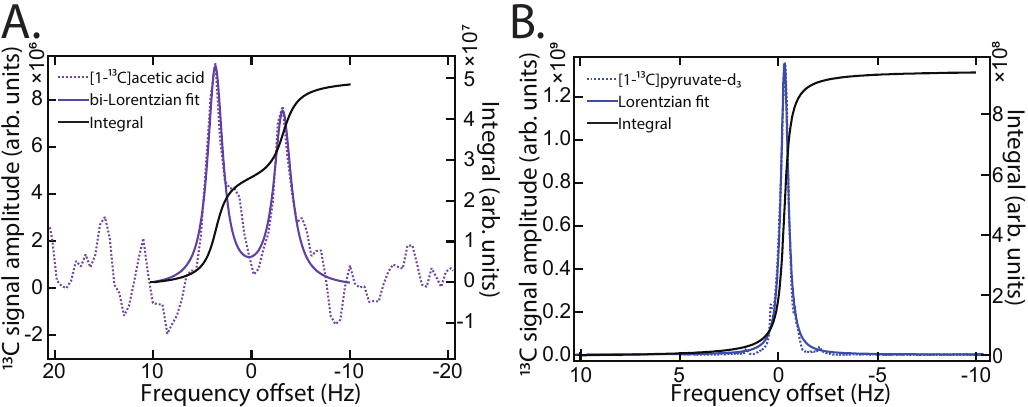}
\caption{\textbf{Spectra for calculating an estimate of $^{13}$C polarization at 6.5~mT. A.} [1-$^{13}$C] acetic acid spectrum acquired with 256 averages of a 45\degree~pulse every 15~s applied to a 31.4~mmol sample. A fit with the sum of two Lorentzians $y=a/((x-x_0)^2+b)+c/((x-x_0)^2+d)$ is plotted as a solid purple line and its integral plotted in black. \textbf{B.} SLIC SABRE hyperpolarized [1-$^{13}$C]pyruvate-d$_3$ spectrum acquired in a single acquisition following 120~s SLIC SABRE with 24~sccm pH$_2$ flow through a sample containing 11.4~µmol pyruvate. A Lorentzian fit of the form $y=a/((x-x_0)^2+b)$ is plotted as a solid blue line and its integral plotted in black.}
\label{Fig_13CPolCal}
\end{center}
\end{figure}
%%%%%%%%%%%%%%%%%%%%%%%%%%%%%%%%%

\section*{Supplementary Section 2: SLIC optimization}
%%%%%%%%%%%%%%%%%%%%%%%%%%%%%%%%%%%%%%%%%%%%%%%%%%%%%%%%%%%%%%%%%%%%%%%%%%%%%%%%%%%%%%%%%%%
%%%%%%%%%%%%%%%%%%%%%%%%%%%%%%%%%%%%%%%%%%%%%%%%%%%%%%%%%%%%%%%%%%%%%%%%%%%%%%%%%%%%%%%%%%%
%%%%%%%%%%%%%%%%%%%%%%%%%%%%%%%%%%%%%%%%%%%%%%%%%%%%%%%%%%%%%%%%%%%%%%%%%%%%%%%%%%%%%%%%%%%
%%%%%   SLIC Optimization FIGURE    %%%%%
\begin{figure}
\begin{center}
\includegraphics[width=16cm]{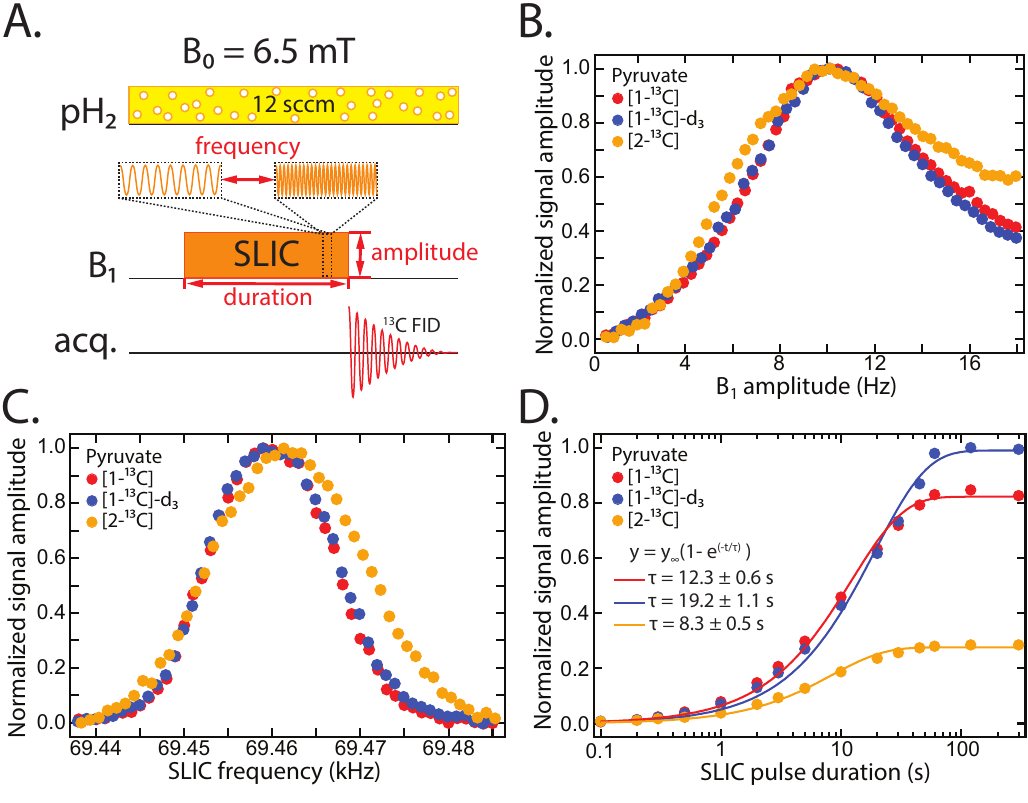}
\caption{\textbf{Optimization of spin-lock properties for SLIC SABRE of $^{13}$C-enriched pyruvate. A.} Pulse sequence diagram of a SLIC SABRE experiment showing the spin-lock variables investigated independently in B-D. \textbf{B.} Hyperpolarized $^{13}$C signal as a function of spin-lock amplitude for protonated (red) and deuterated (blue) [1-$^{13}$C]pyruvate and [2-$^{13}$C]pyruvate (yellow). The most efficient transfer of spin order from pH$_2$ to $^{13}$C occurs at 10.1±0.3 Hz. \textbf{C.} Transverse magnetization as a function of spin-lock frequency. Varying the RF frequency either side of the $^{13}$C resonance at 69.46~kHz for [1-$^{13}$C]pyruvate and 69.462 kHz for [2-$^{13}$C]pyruvate leads to a reduction in the magnetization generated in the transverse plane. \textbf{D.} Build up of hyperpolarized $^{13}$C magnetization with spin-lock duration. Data are fitted with an exponential function, plotted as a solid line, to quantitatively characterize build up rates. For all plots B-D, each data point is the integral of the pyruvate spectrum acquired following SLIC SABRE with the relevant independent variable altered from a standard experiment template of a 10 s spin-lock pulse on resonance, 10~Hz $B_1$ amplitude, 12 sccm pH$_2$ flow, and 4.4 ± 0.2\degree C sample temperature. For B and C, data are normalized to the maximum for each trace to aid comparison. For D, data are normalized to the deuterated pyruvate maximum to show the relative polarization achieved for each sample.}
\label{Fig_SLICoptimization}
\end{center}
\end{figure}
%%%%%%%%%%%%%%%%%%%%%%%%%%%%%%%%%
In this supplementary section, we turn to optimizing the parameters of the spin-lock pulse driving SLIC SABRE. The key advantages of performing in situ SABRE hyperpolarization and detection are: Firstly, that there is no need to transfer samples from an external polarizer to the MRI scanner and secondly, that the reversible chemical exchange in SABRE enables many experiments over time with the same sample. Experiments are reliably repeatable many times over using a single sample and negligible relaxation occurs between polarization and signal acquisition. We investigated how altering the frequency, amplitude and duration of the spin-lock affected the $^{13}$C polarization achieved following SLIC SABRE as shown in Supplementary Figure~\ref{Fig_SLICoptimization}A.

Supplementary Figure~\ref{Fig_SLICoptimization}B shows the effect of tuning the amplitude of the spin-lock. The x-axis is plotted in terms of frequency experienced by spins in the rotating frame to allow for easy comparison to the relevant J-coupling, quoted in Hz rather than in µT of $B_1$ field strength as is more common for SABRE SHEATH. For the calibration method required to calculate the $B_1$ amplitude of a given output power from the spectrometer, refer to Supplementary Section 3. We find that the maximum occurs at $10.1\pm0.3$~Hz. Previously measured values for the J-coupling between hydride groups and the carbon in the 1st position on pyruvate bound to a SABRE catalyst molecule are $J_{HH}=-10.48$~Hz, $J_{HC}=0.55,~0.014$~Hz \cite{Assaf2024} The combined hydride-carbon cross-coupling in the complex compares very closely with our measured optimum $B_1$ amplitude for SLIC SABRE. Similarly to the results for the SLIC frequency sweep, we note that the range of $B_1$ amplitudes that produce $^{13}$C hyperpolarization is relatively broad, similarly explained by the efficiency of SLIC for magnetization transfer where the frequency difference between spin species is large. The [2-$^{13}$C]pyruvate trace is slightly broader than for [1-$^{13}$C]pyruvate. We attribute this to a combination of 2 effects: Firstly, the stronger J-coupling of the carbon in the 2 position to the methyl group alters the J-coupling network through which polarization flows from the hydrides on the SABRE catalyst. Secondly, after 10~s SLIC SABRE [2-$^{13}$C]pyruvate has reached a higher fraction of its saturation polarization relative to [1-$^{13}$C]pyruvate. This flattens the profile of the SLIC-amplitude vs enhancement curve as saturation effects begin to depress the maximum. This flattening effect occurs for [1-$^{13}$C]pyruvate when equivalent experiments are run with longer SLIC pulses. %Show in supp? Or don't bother talking about this at all?

%I'm very unsure if I've explained the above correctly! Not sure if breaking the singlet state is sufficient to drive polarization to 13C or if the SLIC amplitude should match the sum/difference of J_HH±J_HC?%

Supplementary Figure \ref{Fig_SLICoptimization}C plots how the $^{13}$C signal acquired immediately following SLIC SABRE varies with the frequency of the spin-lock pulse. The maximum $^{13}$C signal is acquired with the spin-lock on resonance with the $^{13}$C Larmor frequency. The maxima occur at 69.460~kHz and 69.462~kHz for [1-$^{13}$C]pyruvate and [2-$^{13}$C]pyruvate respectively. The difference being due to chemical shift. As the SLIC frequency is varied either side of resonance the magnetization generated in the transverse plane decreases. If the spin-lock is only a few Hz off resonance, magnetization is also generated along the longitudinal axis.\cite{Pravdivtsev2023} While this process is less efficient, magnetization generated in this fashion can have the advantage that it decays with $T_1$ rather than $T_{1\rho}$ or $T_2$. This has been found to be a necessary method for generating bulk hyperpolarization in solution in cases where the chemical shift between free and bound pyruvate species means the bandwidth of the spin lock is insufficient to affect free pyruvate. One of the great advantages of performing SLIC SABRE at low field is that there is little chemical shift between bound and free pyruvate species. We note that the range of SLIC frequencies over which transverse magnetization is generated is much greater than the both the few-Hz linewidth of the $^{13}$C spectra and the sub-Hz spectral width of the 10~s spin-lock pulses. This occurs because SLIC efficiency is determined by the frequency difference between magnetically inequivalent spins.\cite{DeVience2013} Here, the comparatively large difference between the carbon spin in pyruvate and the proton singlet state on the SABRE catalyst lead to a SLIC SABRE hyperpolarization process that is robust to frequency offsets $\pm10$~Hz either side of the $^{13}$C Larmor frequency.

The last SLIC pulse property to address is spin-lock duration, shown in Supplementary Figure \ref{Fig_SLICoptimization}D. As the spin-lock is applied, magnetization builds up in the transverse plane until it saturates at some level determined by the rate of pH$_2$ bubbling, chemical exchange, efficiency of polarization transfer and $T_{1\rho}$ relaxation. As discussed, SLIC polarization transfer is very efficient in this regime, and the pH$_2$ bubbling rate for these measurements was sufficient not to unduly restrict polarization, so the dominant processes determining the build up rate in this case are the pyruvate $T_1$ and chemical exchange rates. We find that for an equal pH$_2$ bubbling rate, deuterated pyruvate builds up to the highest polarization level while [2-$^{13}$C]pyruvate polarization is comparatively modest. We attribute this difference to the longer $T_1$ of the deuterated pyruvate, allowing for a higher $^{13}$C polarization to accumulate in the bulk solution before $T_1$ relaxation balances SLIC SABRE hyperpolarization. We draw the reader's attention to the speed of the hyperpolarization process in SLIC SABRE, emphasizing that these results are achieved close to room temperature and result in highly polarized samples in a matter of seconds. Compared to dissolution-DNP procedures this hyperpolarization process is more than an order of magnitude faster.
% For 2-$^{13}$C pyruvate, the greater splitting from J-coupling may also hamper polarization transfer efficiency when the narrow bandwidth spin-lock pulse is parked at one frequency. Frequency multiplexing? Modulation schemes? Deuteration?

%%%%%%%%%%%%%%%%%%%%%%%%%%%%%%%%%%%%%%%%%%%%%%%%%%%%%%%%%%%%%%%%%%%%%%%%%%%%%%%%%%%%%%%%%%%
%%%%%%%%%%%%%%%%%%%%%%%%%%%%%%%%%%%%%%%%%%%%%%%%%%%%%%%%%%%%%%%%%%%%%%%%%%%%%%%%%%%%%%%%%%%
%%%%%%%%%%%%%%%%%%%%%%%%%%%%%%%%%%%%%%%%%%%%%%%%%%%%%%%%%%%%%%%%%%%%%%%%%%%%%%%%%%%%%%%%%%%
\section*{Supplementary Section 3: Pulse calibration and spin relaxation times}
%%%%%%%%%%%%%%%%%%%%%%%%%%%%%%%%%%%%%%%%%%%%%%%%%%%%%%%%%%%%%%%%%%%%%%%%%%%%%%%%%%%%%%%%%%%
%%%%%%%%%%%%%%%%%%%%%%%%%%%%%%%%%%%%%%%%%%%%%%%%%%%%%%%%%%%%%%%%%%%%%%%%%%%%%%%%%%%%%%%%%%%
%%%%%%%%%%%%%%%%%%%%%%%%%%%%%%%%%%%%%%%%%%%%%%%%%%%%%%%%%%%%%%%%%%%%%%%%%%%%%%%%%%%%%%%%%%%
In this supplementary section we explain how we calibrate our $B_1$ amplitude before moving on to show results from experiments measuring spin-lattice ($T_1$) and spin-spin ($T_2$) relaxation times. To calibrate spin-lock amplitude, the quantitative relationship between nutation frequency and RF pulse power was determined in two steps: First, a rough calibration was performed by ramping the electromagnet down to 1.6~mT to bring protons on resonance at 69.5~kHz. Measuring a high quality Rabi nutation curve on a CuSO$_4$-doped water sample at this field is readily achievable within 10 minutes in our system with 16 averages of a simple pulse-acquire sequence for a 1~ms pulse length and a series of pulse amplitudes. Fitting the resulting curve with a decaying sinusoid function provides the pulse power calibration for protons, which can be used to calculate the corresponding values for $^{13}$C at 6.5~mT by increasing the pulse power by the square of the ratio of the gyromagnetic ratios of $^1$H and $^{13}$C (12dB). This first calibration enabled spin-lock pulses of appropriate amplitude to be applied to SABRE samples to hyperpolarize pyruvate in situ via SLIC SABRE, giving sufficient $^{13}$C signal at 6.5~mT for further calibration. 
The second calibration step was performed by measuring a nutation curve created by applying a hard RF pulse immediately following a spin-lock pulse that generated net $^{13}$C magnetization in the transverse plane as shown in Supplementary Figure~\ref{Fig_B1cal}A. The resulting nutation curve, plotted in Supplementary Figure~\ref{Fig_B1cal}B, is in the functional form of a cosine as the hard pulse tips the net magnetization vector away from the transverse plane. Extracting the nutation frequency allows for the calculation of the strength of $B_1$ and calibration of pulse amplitudes. For example, a 90\degree~ pulse length of 1~ms, corresponds to a full 360\degree~ rotation in 4~ms or 250~Hz with 450~µW output from the console. From that calibration point, increasing the spectrometer output attenuation by 27~dB reduced the effective $B_1$ frequency to 11.1~Hz, and from there the transmit amplitude was adjusted to perform $B_1$ amplitude sweeps such as those shown in Supplementary Figure~\ref{Fig_SLICoptimization}B. We note that this calibration method assumes the spectrometer transmit amplitude adjustment is linear and the attenuation values are accurate. 
%%%%%   B1 calibration FIGURE   %%%%%
\begin{figure} %%% "figure" for single-column width figure %%%
\begin{center}
\includegraphics[width=16cm]{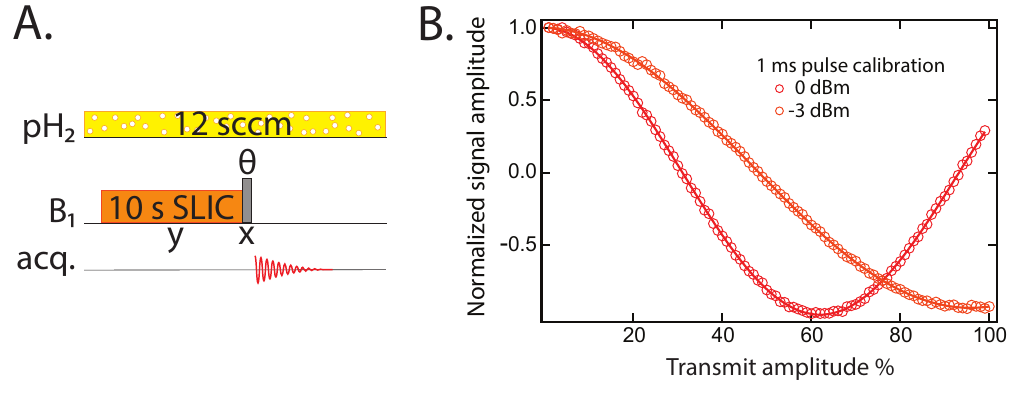}
\caption{\textbf{Pulse calibration with magnetization generated by SLIC SABRE. A.} Pulse sequence for secondary tip-angle calibration. Following the application of SLIC SABRE to generate transverse $^{13}$C magnetization in a pyruvate SABRE solution, a hard pulse $\theta$ is applied with orthogonal phase to the spin-lock pulse, rotating magnetization out of the transverse plane. \textbf{B.} Repeating the pulse sequence in A, varying the spectrometer transmit amplitude for 1~ms pulses with output power of 0~dBm (red) or -3~dBm (orange), results in nutation curves with the functional form of a decaying cosine (plotted as solid lines). The transmit amplitude corresponding to a 90\degree~ pulse occurs at the first zero crossing.}
\label{Fig_B1cal}
\end{center}
\end{figure}
%%%%%%%%%%%%%%%%%%%%%%%%%%%%%%%%%

Having demonstrated that we have reliable tip angle calibration for hard pulses we move on to the investigation of spin-lattice relaxation. For in vivo hyperpolarized imaging applications, spin-lattice relaxation time ($T_1$) is critical because it determines how long the hyperpolarized signal remains visible and how far down the chain of metabolic products information can be extracted. Much of the success of pyruvate as a hyperpolarized metabolite is due to its comparatively long $T_1$ at tesla field strengths: [1-$^{13}$C]pyruvate and [1-$^{13}$C]pyruvate-d$_3$ $T_1$ times are approximately 2 and 4 minutes at 2~T.\cite{Schmidt2023} 

Pyruvate $T_1$ decay curves measured at 6.5~mT are shown in Supplementary Figure \ref{Fig_T1}. Our pulse sequence for measuring $T_1$ in a SLIC SABRE experiment is drawn in Supplementary Figure \ref{Fig_T1}A. The typical SLIC SABRE pulse sequence was modified by the addition of 2 hard 90\degree~ pulses. As SLIC SABRE on resonance produces net magnetization aligned with $B_1$ in the transverse plane of the Bloch sphere, to measure $T_1$ relaxation the magnetization is rotated to lie along the longitudinal axis by the first 90\degree~ pulse. After some evolution time, the magnetization is then returned to the transverse plane for readout by the application of another 90\degree~ pulse with the opposite RF phase and an FID acquired. The resulting decay curves were fitted with a decaying exponential of the form \(y=e^{-t/T_1}\) where \(t\) is the time between the hard 90\degree~ pulses and \(T_1\) is the spin-lattice relaxation time.

The resulting data are plotted in Supplementary Figure \ref{Fig_T1}B and fitted to a decaying exponential function to extract the characteristic $T_1$ decay constant. We measure $T_1$ values of $34.0\pm0.8$~s, $20.7\pm1.3$~s and $45.0\pm1.3$~s for [1-$^{13}$C]pyruvate, [2-$^{13}$C]pyruvate and [1-$^{13}$C]pyruvate-d$_3$ respectively. These relaxation times are consistent with an interpolation from previous results measured at microtesla and tesla fields.\cite{Schmidt2023} We note that these times are measured while the sample is undergoing constant pH$_2$ bubbling action that continues during the delay in our $T_1$ measurement sequence. This does not appear to affect $T_1$ relaxation as our results are comparable to experiments using small tip angle sequences on pyruvate samples hyperpolarized outside the 6.5~mT scanner and transferred inside for measurement, shown in Supplementary Figure \ref{Fig_SmallTip}. However, there is reason to believe that given there are ongoing SABRE processes occurring, including proton hyperpolarization at 6.5~mT, our $T_1$ values measured here should be regarded as the "effective $T_1$" specific to these conditions. 

%%%%%   T1 DECAY FIGURE    %%%%%
\begin{figure} %%% "figure" for single-column width figure %%%
\begin{center}
\includegraphics[width=16cm]{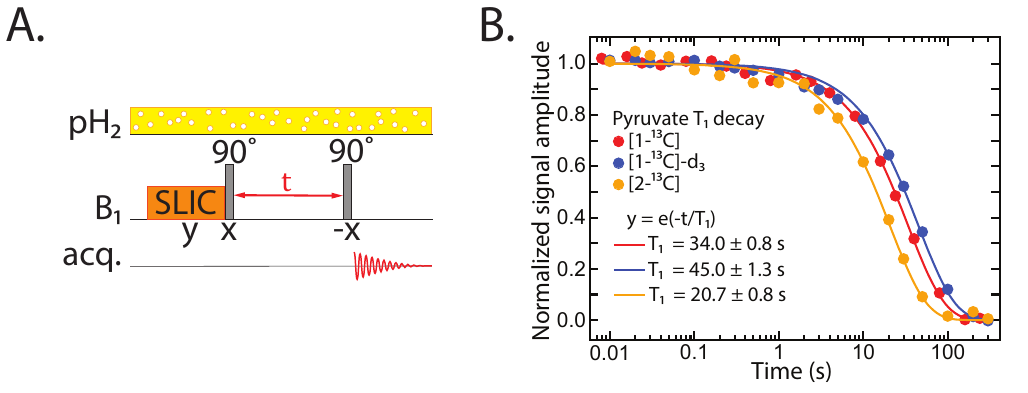}
\caption{\textbf{Spin-lattice relaxation of hyperpolarized [1-$^{13}$C]pyruvate. A.} Pulse sequence for measuring $T_1$ decay of SLIC SABRE hyperpolarized pyruvate. Following 10~s SLIC SABRE on resonance at 69.46~kHz, at 10~Hz $B_1$ amplitude, $^{13}$C magnetization was rotated out of the transverse plane to lie along the longitudinal axis by a hard 90° pulse, before being returned to the transverse plane by another hard 90° to measure $T_1$ relaxation of the hyperpolarized state. pH$_2$ flow was held constant at 12 sccm and temperature was held at 4.4~±~0.2\degree C. \textbf{B.} Measured hyperpolarization decay curves. Data points are integrals of [1-$^{13}$C]pyruvate (red), [2-$^{13}$C]pyruvate (yellow) and [1-$^{13}$C]pyruvate-d$_3$ (blue) spectra. Solid lines are exponential fits.}
\label{Fig_T1}
\end{center}
\end{figure}
%%%%%%%%%%%%%%%%%%%%%%%%%%%%%%%%%

%%%%%   Transfer and small tip decay FIGURE    %%%%%
\begin{figure} %%% "figure" for single-column width figure %%%
\begin{center}
\includegraphics{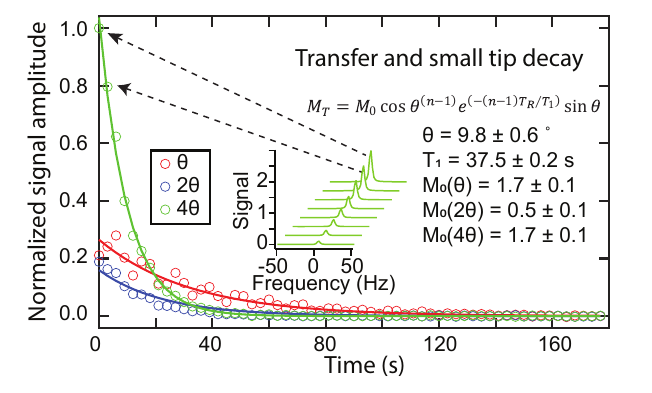}
\caption{\textbf{Small tip angle measurements of hyperpolarization decay.} [1-$^{13}$C]pyruvate hyperpolarized outside the 6.5~mT scanner with a SLIC SABRE polarizer (described in reference \cite{McBride2025}) and transferred to 6.5~mT for measurement. Following shuttling of a hyperpolarized 
$^{13}$C sample into the scanner, a pulse sequence of 60 RF pulses, spaced 3~s apart, was applied and the FID acquired with a solenoid coil tuned and matched to 69.5kHz. For subsequent experiments the tip angle was incremented by doubling the pulse amplitude. As the 3 small tip angle decay curves then have a common $T_1$ and known relationship between their applied tip angles, they can be simultaneously fitted to 3 variations of the equation $M_T=M_0cos\theta^{(n-1)}e^{(-(n-1)T_R/T_1)}sin\theta$, where $M_T$ is the transverse magnetization acquired following the $n$th tip, $M_0$ is the initial magnetization delivered for each run, $\theta$ is the tip angle, and $T_R$ is the repetition time of 3~s.}
\label{Fig_SmallTip}
\end{center}
\end{figure}
%%%%%%%%%%%%%%%%%%%%%%%%%%%%%%%%%

With a measure of $T_1$ established for our system, we consider $T_2$ relaxation. The variable tip angle schedule used by our SLIC-shot bSSFP sequence is determined by the choice of initial tip, the number of pulses, $T_1$, $T_R$ and $T_2$.\cite{Deppe2012} To determine an appropriate $T_2$ value for calculating our tip angle schedule we used a sequence similar to a Hahn echo, as shown in Supplementary Figure \ref{Fig_T2}A. The typical Hahn echo experiment was modified by the removal of the initial 90\degree~ pulse as the SLIC SABRE spin-lock generates transverse magnetization. Bubbling was continued throughout, matching our other measurements. Results from this Hahn-echo-style measurement performed on [1-$^{13}$C]pyruvate-d$_3$ are plotted in Supplementary Figure \ref{Fig_T2}B and fitted with an exponential decay of the form \(y=e^{-t/T_2}\) where \(t\) is the time between the end of the spin-lock and the center of the spin echo, and \(T_2\) is the spin-spin relaxation time. The mono exponential decay appears insufficient to describe all the features of the observed behavior, but gives a good representative value to inform our variable tip angle schedule. Given the echo time used in our SLIC-shot MRI experiments was 54~ms, much less than our $T_2$ of $2.15\pm0.15$s, our variable tip angle schedule was relatively insensitive to $T_2$ compared to $T_1$ and our choices of initial tip angle and number of lines of k-space to acquire.
%%%%%   T2 DECAY FIGURE    %%%%%
\begin{figure} %%% "figure" for single-column width figure %%%
\begin{center}
\includegraphics[width=16cm]{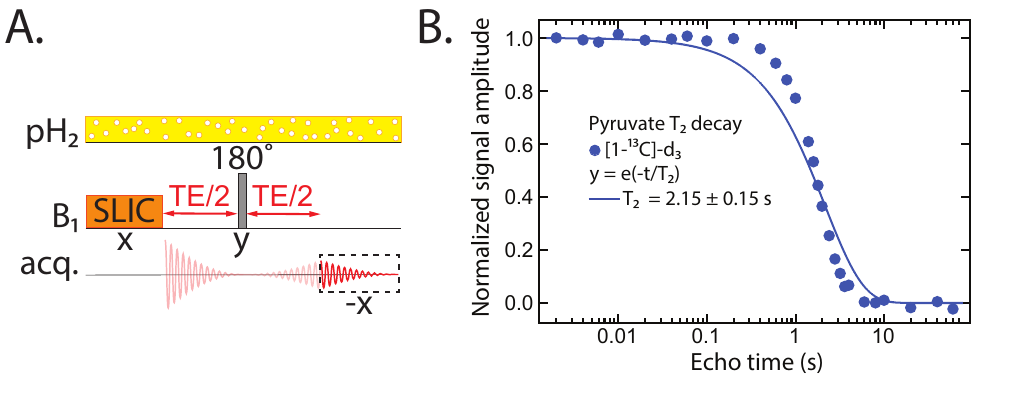}
\caption{\textbf{Spin-spin relaxation of hyperpolarized [1-$^{13}$C]pyruvate-d$_3$. A.} Pulse sequence for measuring $T_2$ decay of SLIC SABRE hyperpolarized pyruvate. Following 30~s SLIC SABRE on resonance, $^{13}$C magnetization dephases in the transverse plane before a 180\degree~pulse is applied with orthogonal phase to the initial spin-lock to refocus the magnetization into a spin echo. To keep the acquisition length consistent across all echo times only the second half of the spin echo is acquired. pH$_2$ flow was held constant at 30 sccm and temperature was held at 4.4~±~0.2\degree C. \textbf{B.} Measured spin echo decay curves. Data points are integrals of [1-$^{13}$C]pyruvate-d$_3$ spectra, the solid line is an exponential fit.}
\label{Fig_T2}
\end{center}
\end{figure}
%%%%%%%%%%%%%%%%%%%%%%%%%%%%%%%%%

%%%%%%%%%%%%%%%%%%%%%%%%%%%%%%%%%%%%%%%%%%%%%%%%%%%%%%%%%%%%%%%%%%%%%%%%%%%%%%%%%%%%%%%%%%%
%%%%%%%%%%%%%%%%%%%%%%%%%%%%%%%%%%%%%%%%%%%%%%%%%%%%%%%%%%%%%%%%%%%%%%%%%%%%%%%%%%%%%%%%%%%
%%%%%%%%%%%%%%%%%%%%%%%%%%%%%%%%%%%%%%%%%%%%%%%%%%%%%%%%%%%%%%%%%%%%%%%%%%%%%%%%%%%%%%%%%%%
\section*{Supplementary Section 4: Further discussion of fine features in pyruvate spectra}
%%%%%%%%%%%%%%%%%%%%%%%%%%%%%%%%%%%%%%%%%%%%%%%%%%%%%%%%%%%%%%%%%%%%%%%%%%%%%%%%%%%%%%%%%%%
%%%%%%%%%%%%%%%%%%%%%%%%%%%%%%%%%%%%%%%%%%%%%%%%%%%%%%%%%%%%%%%%%%%%%%%%%%%%%%%%%%%%%%%%%%%
%%%%%%%%%%%%%%%%%%%%%%%%%%%%%%%%%%%%%%%%%%%%%%%%%%%%%%%%%%%%%%%%%%%%%%%%%%%%%%%%%%%%%%%%%%%

% FROM THE MAIN TEXT:
%We attribute the fine details in our 1-$^{13}$C pyruvate spectra to imperfect enrichment and/or exchange of protons and deuterons occurring over time.\cite{Barskiy2014} To investigate other potential explanations we performed additional NMR spectroscopy at 11.7~T on SABRE solutions after measurement at ULF, detailed in Supplementary section \ref{}.

The fine details in our [1-$^{13}$C]pyruvate and [1-$^{13}$C]pyruvate-d$_3$ spectra shown in Figure 4 of the main text include peaks in the wings of the deuterated pyruvate spectrum and a peak in the middle of the quartet in the protonated pyruvate spectrum. Extending the discussion of spectral features in the main text, here we explain why we attribute their origin to the exchange of protons and deuterons, and describe how these additional enigmatic features evolve over time in SABRE solutions. While secondary to the main narrative of our manuscript we share these measurements as useful validation of the the need to prepare a fresh batch of sample for each SABRE experiment. They are also of interest as they reinforce evidence that SABRE cannot always be rigorously considered a non-hydrogenative parahydrogen induced polarization process.\cite{Barskiy2014} Further, to investigate other potential explanations we show 11.7~T NMR spectroscopy data from SABRE solutions after their use in SLIC SABRE experiments at 6.5~mT.

For the simple case of [1-$^{13}$C]pyruvate-d$_3$, the quartet of peaks about the primary signal are most likely due to imperfect deuteration of the methyl group. The quartet of peaks matches the [1-$^{13}$C]pyruvate quartet and the provider-quoted deuterium fraction is $\geq$97\%. While this explanation is straightforward, it also offers a clue to help explain our observations for protonated pyruvate: The unexpected central feature in the [1-$^{13}$C]pyruvate spectrum occurs at the same frequency as the main [1-$^{13}$C]pyruvate-d$_3$ peak, suggesting a common origin. This central feature also increases in magnitude over time, as shown in Supplementary Figure \ref{Fig_SpectraOverTime}, which is consistent with deuteration occurring, mediated by the SABRE catalyst, with the methanol-d$_4$ solvent providing an abundant source of deuterium. 

%%%%%   Spectra over time FIGURE    %%%%%
\begin{figure} %%% "figure" for single-column width figure %%%
\begin{center}
\includegraphics{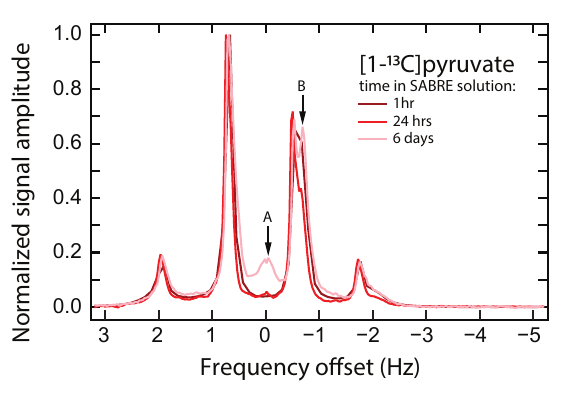}
\caption{\textbf{[1-$^{13}$C]pyruvate $^{13}$C spectral evolution.} A single batch of SABRE solution was prepared, stored refrigerated under argon, and sampled for SLIC SABRE experiments at 6.5~mT at 3 time points. Arrows draw attention to features at "A" and "B" that are not explained by the quartet expected from [1-$^{13}$C]pyruvate. We attribute feature A to deuterated pyruvate, and feature B to free [2-$^{13}$C]pyruvate, likely present in small concentrations from imperfect enrichment.}
\label{Fig_SpectraOverTime}
\end{center}
\end{figure}
%%%%%%%%%%%%%%%%%%%%%%%%%%%%%%%%%

High-field NMR spectra measured at 11.7~T are shown in Supplementary Figure \ref{Fig_HighField}, comparing [1-$^{13}$C]pyruvate and [1-$^{13}$C]pyruvate-d$_3$ $^1$H and $^{13}$C spectra from samples following SLIC SABRE experiments. These samples were stored at room temperature for several hours, likely accelerating any processes at work in the data shown in Supplementary Figure \ref{Fig_SpectraOverTime}. No significant variation can be identified between samples. There are no spectral features sufficiently near the pyruvate peak to suggest an impurity may account for our observations at 6.5~mT. At 6.5~mT, the peak at about 176 ppm in the high-field $^{13}$C spectra would be seen in the hyperpolarized 6.5~mT spectrum +1.7 Hz away from pyruvate, and the peak around 181 ppm would be at about +3 Hz. Neither peak appears in the SLIC-SABRE hyperpolarized spectrum, suggesting these species are not hyperpolarized under our SLIC SABRE conditions.

%%%%%   High field spectra FIGURE    %%%%%
\begin{figure} %%% "figure" for single-column width figure %%%
\begin{center}
\includegraphics[width=16cm]{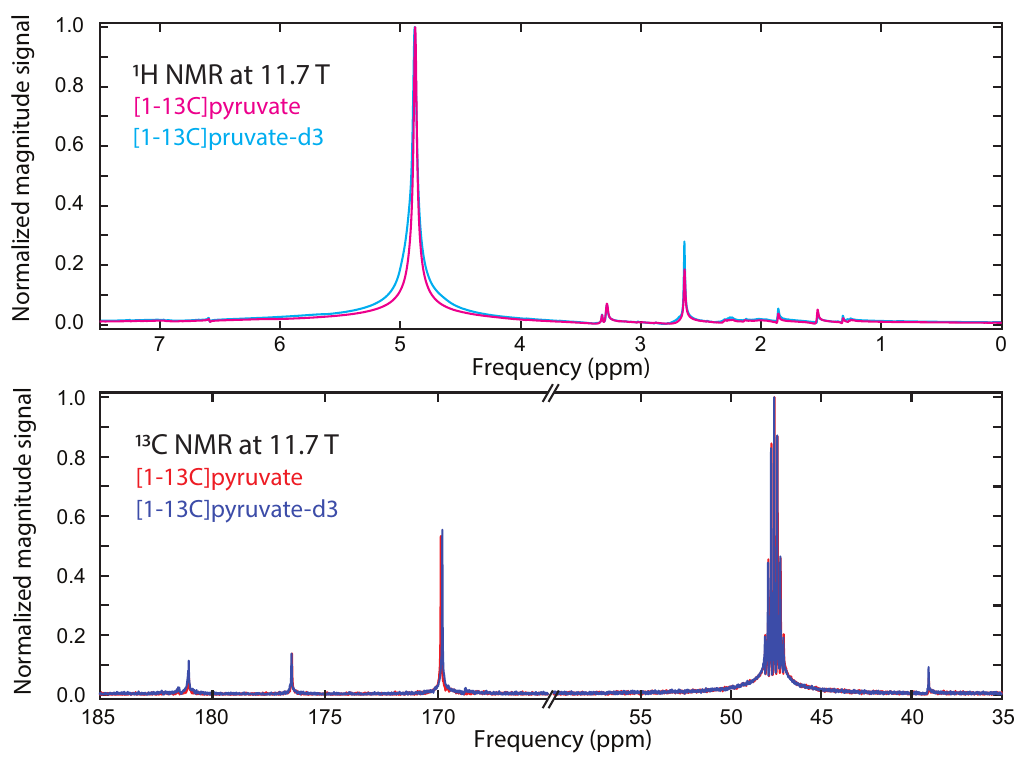}
\caption{\textbf{NMR spectra of pyruvate SABRE solutions at 11.7~T following SLIC SABRE experiments at 6.5~mT. Top:} $^1$H NMR of [1-$^{13}$C]pyruvate (magenta) and [1-$^{13}$C]pyruvate-d$_3$ (cyan). $^1$H spectra are dominated by a broad hydroxl group feature. \textbf{Bottom:} $^{13}$C NMR of [1-$^{13}$C]pyruvate (red) and [1-$^{13}$C]pyruvate-d$_3$ (blue). Methanol-d$_4$ accounts for the largest feature -- a septet from the deuterated methyl group -- with the second largest feature around 170~ppm the [1-$^{13}$C]pyruvate peak. We attribute the feature at approximately 176~ppm to pyruvate hydrate and suggest the small peaks around 181~ppm may be zymonic acid or parapyruvate.\cite{Harris2018} Spectra were measured with a JEOL 500~MHz NMR console with the frequency axis shown here calculated from the spectrometer offset calibration rather than an independent reference.}
\label{Fig_HighField}
\end{center}
\end{figure}
%%%%%%%%%%%%%%%%%%%%%%%%%%%%%%%%%

%%%%%%%%%%%%%%%%%%%%%%%%%%%%%%%%%%%%%%%%%%%%%%%%%%%%%%%%%%%%%%%%%%%%%%%%%%%%%%%%%%%%%%%%%%%
%%%%%%%%%%%%%%%%%%%%%%%%%%%%%%%%%%%%%%%%%%%%%%%%%%%%%%%%%%%%%%%%%%%%%%%%%%%%%%%%%%%%%%%%%%%
%%%%%%%%%%%%%%%%%%%%%%%%%%%%%%%%%%%%%%%%%%%%%%%%%%%%%%%%%%%%%%%%%%%%%%%%%%%%%%%%%%%%%%%%%%%
\section*{Supplementary Section 5: Optimizing conditions for SABRE }
%%%%%%%%%%%%%%%%%%%%%%%%%%%%%%%%%%%%%%%%%%%%%%%%%%%%%%%%%%%%%%%%%%%%%%%%%%%%%%%%%%%%%%%%%%%
%%%%%%%%%%%%%%%%%%%%%%%%%%%%%%%%%%%%%%%%%%%%%%%%%%%%%%%%%%%%%%%%%%%%%%%%%%%%%%%%%%%%%%%%%%%
%%%%%%%%%%%%%%%%%%%%%%%%%%%%%%%%%%%%%%%%%%%%%%%%%%%%%%%%%%%%%%%%%%%%%%%%%%%%%%%%%%%%%%%%%%%

Our final supplementary section considers the optimum pH$_2$ flow rate and temperature conditions for maximizing pyruvate polarization at 6.5~mT. These results are most important in the context of informing future polarizer and reactor design. For completeness, we also briefly present our pH2 generator and describe how our gas handling and temperature control circuits interface with the SLIC SABRE experiment situated inside the 6.5~mT scanner.

The gas handling circuit used to supply pH$_2$ to the SABRE solution inside the RF-shielded scan room is shown in Supplementary Figure~\ref{Fig_ExpSetup2}A. Before an experiment begins the gas lines are pumped and flushed with pH$_2$. Bypass valves built into the circuit aid this process and enable pH2 bottle and SABRE sample swaps without necessitating depressurizing all gas lines. A purge valve inside the scan room acts as a test point to confirm hydrogen delivery and helps leak checking by providing a true positive reading location. To bubble pH$_2$ through the SABRE solution the pressure regulator on the pH$_2$ supply bottle is set to 110~psi and the back pressure regulator set to 95~psi. The mass flow controller (MFC) can then throttle pH$_2$ independently.

% pH2 flow varied
For experiments investigating the effect of varying pH$_2$ flow rate, SLIC pulses were applied on resonance, with 10~Hz amplitude and 30~s duration. After each adjustment of the MFC flow setting, multiple consecutive SLIC SABRE experiments were performed to verify the polarization achieved had reached repeatable equilibrium at the new flow rate. This usually occurred within the first 2 30~s repetitions. Once polarization levels had stabilized at a new flow setting, results were averaged and the standard deviation used to quantify the uncertainty. Build up curves were fitted with an exponential of the form \(y=1-e^{-x/\tau}\) where $x$ was the pH$_2$ flow rate and \(\tau\) the build up constant.

% Temperature dependence
To investigate the temperature dependence of SLIC SABRE at 6.5~mT the same basic experiment for investigating pH$_2$ flow rate was employed, but instead of varying MFC setting, the chiller setpoint was changed and the sample temperature allowed to re-equilibrate before each set of measurements, while pH$_2$ was held constant at 12~sccm. For each setpoint, the temperature was stable to within ±0.2\degree C, and 4 averages were acquired with 30~s SLIC on resonance. The standard deviation between each set of four measurements was used to estimate uncertainty.

Results plotting the signal enhancement as a function of pH$_2$ flow rate are presented in Supplementary Figure~\ref{Fig_ExpSetup2}B. Each point plots the $^{13}$C signal acquired immediately following 30~s SLIC SABRE on resonance. Exponential curves are plotted as a guide to the build up rate rather than to represent an analytical solution to the competing processes that govern these experiments. The behavior here is dominated by the fluid dynamics of bubbling through a small capillary into the bottom of a narrow tube. With ideal uniform diffusion of pH$_2$ through the solution, saturation would occur when pH$_2$ is supplied fast enough to ensure every H$_2$ that binds to a catalyst molecule is statistically likely to be fresh pH$_2$ rather than oH$_2$. However, in our the case saturation occurs when bubbling forces the sample out of the NMR coil. This effect is already pronounced at 24~sccm and has been foreshadowed by the $^{13}$C MRI results in Figure \ref{Fig_multiSLICshotMRI} showing hyperpolarized solution being pushed upwards. This highlights the need for optimized polarizer designs and the benefit of reactors that control the distribution of hydrogen gas into solution.\cite{Kempf2024,TomHon2021}

%%%%%   EXPERIMENTAL SETUP #2 FIGURE    %%%%%
\begin{figure} %%% "figure" for single-column width figure %%%
\begin{center}
\includegraphics[width=16cm]{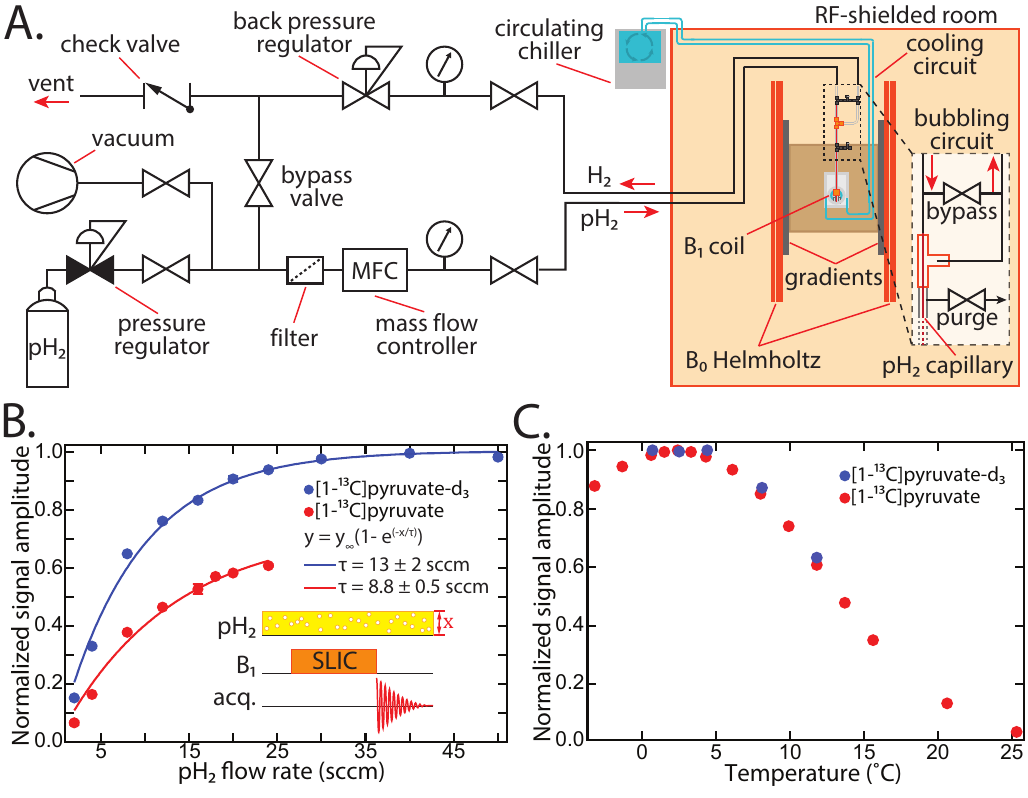}
\caption{\textbf{Parahydrogen and temperature control A.} Schematic representation of the gas handling and cooling circuits used to perform SLIC SABRE experiments inside the 6.5~mT MRI scanner. \textbf{B.} $^{13}$C hyperpolarization build up vs pH$_2$ flow rate for deuterated (blue) and protonated (red) [1-$^{13}$C]pyruvate at 6.5~mT. Insert shows simplified pulse sequence emphasizing the variation in the rate of pH$_2$ bubbling characterized here. Data are normalized by division by the maximum signal from the deuterated pyruvate. Data are fitted with an exponential function, plotted as a solid line, to quantitatively characterize build up rates. Each data point is the integral of four averages of the pyruvate spectrum acquired following 30~s SLIC SABRE on resonance at 69.46~kHz, 10~Hz $B_1$ amplitude, with sample temperature held at at 4.4~±~0.2\degree C and pH$_2$ pressure at 95~psi. Error bars are the standard deviation, plotted when they are larger than the size of the data point. \textbf{C.} Hyperpolarized $^{13}$C signal as a function of temperature. Data points are the integrals of hyperpolarized spectra, measured with 4 averages of 30~s SLIC on resonance at 12~sccm pH$_2$. Data normalized to maximum signal for each pyruvate sample. Error bars are smaller than data points.}
\label{Fig_ExpSetup2}
\end{center}
\end{figure}
%%%%%%%%%%%%%%%%%%%%%%%%%%%%%%%%%

Supplementary Figure \ref{Fig_ExpSetup2}C plots the SLIC SABRE signal as a function of temperature for 1-$^{13}$C pyruvate. Each data point is 4 averages following 30~s SLIC on resonance with pH$_2$ bubbling at 12~sccm. The protonated and deuterated pyruvate samples exhibit the same temperature dependence, with maximum polarization achieved at $2.5\pm1$\degree C. Varying the temperature of the SABRE solution modulates the chemical exchange rate of pyruvate molecules binding and releasing the catalyst. At low temperature, pyruvate binds too strongly to the SABRE catalyst, slowing chemical exchange and constricting the generation of bulk hyperpolarization of free pyruvate solution. At higher temperatures, pyruvate exchanges too rapidly for polarization transfer to occur mediated through J-coupling between the parahydrogen singlet state and pyruvate across the SABRE catalyst. While our results demonstrate that there is a clear optimum temperature, we observe a broad temperature range of >10\degree C where at least 80\% of the maximum enhancement is produced. This shows that the temperature control requirements for efficient SABRE hyperpolarization in future polarizer designs are less strenuous than those for efficient pH$_2$ delivery, homogeneous $B_0$ and $B_1$ fields and rapid sample purification.

Finally, we briefly share the details our pH$_2$ generator, shown in Supplementary Figure \ref{Fig_pH2Generator}. The pH$_2$ generator consists of a closed cycle cryostat (ARS, DE-204S cold head) driven by a helium compressor (ARS-4HW) and monitored by a temperature controller (Lakeshore, LS335). A scroll pump (Pfeiffer, HiScroll12) serves to evacuate the cryostat vacuum shroud before generator operation and pump contaminants out of the H$_2$ gas handling circuit. The cryostat sample chamber is packed with FeOOH catalyst, sieved to $>45$µm, contained with quartz wool (McMaster-Carr) and filter paper (Whatman, grade 602h) to prevent migration through the H$_2$ circuit. During operation, the cryostat temperature is set to 27~K and 500~psi H$_2$ gas is flowed through the catalyst-packed sample chamber, with flow rate maintained at 2~slm by a mass flow controller (Alicat, MCQ-5SLPM). Aluminium cylinders are pumped and flushed with fresh pH$_2$ before filling to 480~psi. Check and bypass valves act to prevent flow through the generator of any contaminants introduced during cylinder swapping. All hydrogen that is released during operation, via the vacuum pump, vent valve, or 550~psi relief valve, is vented into a fume hood.

%%%%%   pH2 GENERATOR FIGURE    %%%%%
\begin{figure} %%% "figure" for single-column width figure %%%
\begin{center}
\includegraphics[width=16cm]{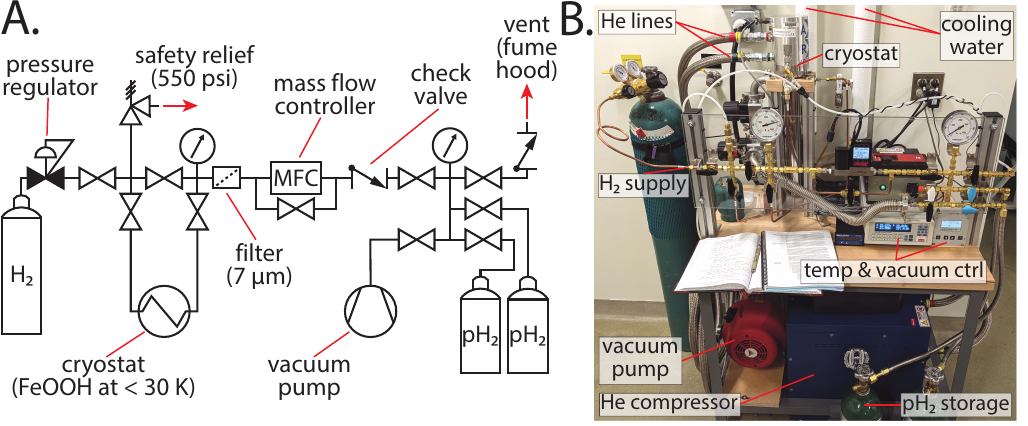}
\caption{\textbf{pH$_2$ generator. A.} H$_2$ gas handling circuit schematic. Only the section of the vacuum system for pumping H$_2$ is shown. \textbf{B.} Labeled photograph of the pH2 generator in operation. The gas handling circuit is mounted on a transparent polycarbonate panel in front of the cryostat, with the compressor, vacuum pump and pH$_2$ cylinders underneath.}
\label{Fig_pH2Generator}
\end{center}
\end{figure}
%%%%%%%%%%%%%%%%%%%%%%%%%%%%%%%%%

\FloatBarrier

%%%%%%%%%%%%%%%%%%%%%%%%%%%%%%%%%%%%%%%%%%%%%%%%%%%%%%%%%%%%%%%%%%%%%%%%%%%%%%%%%%%%%%%%%%%

\clearpage

%\makeatletter 
%\renewcommand{\thefigure}{S\@arabic\c@figure}
%\makeatother

\end{document}